\documentclass[epj]{svjour}
\usepackage{graphics}
\usepackage{subfigure}
\usepackage{float}
\usepackage[T1]{fontenc}
\usepackage{graphicx}
\usepackage{amsmath,amssymb,slashed,bbm,xcolor}
\usepackage{widetext}

\DeclareMathAlphabet{\boldmathe}{T1}{cmr}{bx}{it}

\def\beq{\begin{eqnarray}}
\def\eeq{\end{eqnarray}}
\def\ln{\,\mbox{ln}\,}

\def\Det{\,\mbox{Det}\,}

\def\Tr{\,\mbox{Tr}\,}

\def\al{\alpha}
\def\be{\beta}

\def\ga{\gamma}
\def\de{\delta}
\def\vp{\varepsilon}

\def\ka{\kappa}
\def\la{\lambda}
\def\na{\nabla}

\def\si{\sigma}
\def\om{\omega}

\def\Ga{\Gamma}

\def\La{\Lambda}

\def\B1{ Bianchi-I }
\begin{document}
\sloppy
%this command was addicionated for the text never go over the right margin

%%%%%%%%%%%%%%%%%%%%%%%%%%%%%%%%%
%%%%%%%%%%%%%%%%%%%%%%%%%%%%%%%%%
\title{Renormalization group in super-renormalizable quantum gravity}

%% \subtitle{No subtitle}

\author{Leonardo Modesto\inst{1}
\thanks{\emph{E-mail address:} lmodesto@sustc.edu.cn},
%% , lmodesto1905@icloud.com},
Les\l{}aw Rachwa\l,\inst{1,2}\thanks{\emph{E-mail address:} grzerach@gmail.com},
Ilya L. Shapiro\inst{3,4,5}\thanks{\emph{E-mail address:} shapiro@fisica.ufjf.br}% etc
}                     % Do not remove
\institute{
Department of Physics, Southern University of Science and
Technology, Shenzhen 518055, China
\and \
Instituto de F\'{i}sica, Universidade de Bras\'{i}lia,
70910-900, Bras\'{i}lia, DF, Brazil
\and \
Departamento de F\'{\i}sica, ICE, Universidade Federal de Juiz de Fora,
\\
Campus Universit\'{a}rio - Juiz de Fora, 36036-330, MG, Brazil
\and \
Tomsk State Pedagogical University, Tomsk, 634041, Russia
\and \
Tomsk State University, Tomsk, 634050, Russia}
\date{Received: date / Revised version: date}

\abstract{ One of the main advantages of super-renormalizable
higher derivative quantum gravity models is the possibility to derive
exact beta functions, by making perturbative one-loop calculations.
We perform such a calculation for the Newton constant by using
the Barvinsky-Vilkovisky trace technology. The result is well-defined
in a large class of models of gravity in the sense that the
renormalization group beta functions do not depend on the
gauge-fixing condition. Finally, we discuss the possibility to apply
the results to a large class of nonlocal gravitational theories which
are free of massive ghost-like states at the tree-level.
%%%%%%%%%%%%%%%%%%%%%%%%%%%%%%%%%
%%%%%%%%%%%%%%%%%%%%%%%%%%%%%%%%%
% 11.10.Hi	%%   Renormalization group evolution of parameters
% \and \
% 04.50.Kd}           	%% Modified theories of gravity
%%    PACS code2 \and more}
%%    \subclass{MSC code1 \and MSC code2 \and more}
%% \PACS{
%%     {\emph{MSC:}}{81T16, 81T17, 81T20} \and
%%       {PACS:}{04.62.+v, 11.10.Hi, 11.15.Tk }    } % end of PACS codes
} %end of abstract
\authorrunning{L. Modesto, L. Rachwa\l, I. Shapiro}
\titlerunning{Renormalization group in super-renormalizable quantum gravity}
\maketitle

%%%%%%%%%%%%%%%%%%%%%%%%%%%%%%%%%
%%%%%%%%%%%%%%%%%%%%%%%%%%%%%%%%%
\section{Introduction}
\label{I}

The calculation of quantum corrections always had a very special
role in quantum theories of gravity. The first relevant calculation was
done by t'Hooft and Veltman \cite{hove}, who derived the one-loop
divergences in the quantum version of general relativity, including
coupling to the minimal scalar field. Soon after similar calculations
were performed for gravity-vector and gravity-fermion systems
\cite{dene}. These first calculations have a great merit, regardless
of the fact that the output was shown to be gauge-fixing dependent
\cite{KTT}. Later on one could learn a lot from the two-loop
calculations in general relativity \cite{gosa,ven}. Technically more
complicated are calculations in four-derivative gravity, which
were first performed in \cite{JulTon} and with some corrections in
\cite{frts82}, \cite{avbar86} and finally in \cite{Gauss}, where some
extra control of the calculations was introduced and the hypothesis
of the non-zero effect of the topological Gauss-Bonnet term
\cite{capkim} explored. Let us also mention similar calculations
in the conformal version of the four-derivative theory
\cite{frts82,amm,Weyl}.

The importance of four-derivative quantum gravity is due to its
renormalizability \cite{Stelle77}, which is related to the presence
of massive unphysical ghosts, typical in the higher derivative
field theories. Naturally, there were numerous and interesting
works trying to solve the unitarity problem in this theory
\cite{Tomboulis-77,salstr,antomb}. The mainstream approach is
based on the expectation that the loop corrections may transform
the real massive unphysical pole in the tensor sector of the theory
into a pair of complex conjugate poles, which do not spoil unitarity
within the Lee-Wick quantization scheme \cite{LW}. However,
it was shown that the definite knowledge of whether this scheme
works or not requires an exact non-perturbative beta function
for the coefficient of the Weyl-squared term and for the Newton
constant \cite{Johnston}.  The existing methods to obtain such
a non-perturbative result give some hope \cite{CodPer-06}, but
unfortunately they are not completely reliable\footnote{One of the
reasons is a strong gauge-fixing dependence of the on-shell average
effective action, which was discussed in Yang-Mills theory
\cite{FRG-Lavrov} and is expected to take place also in quantum
gravity.} and, therefore, the situation with unitarity in the
four-derivative quantum gravity is not certain, at least.

Another interesting aspect of quantum corrections in models
of gravity is related to the running of the cosmological constant
$\La_{\rm cc}$ and  especially Newton constant $G$. These
quantum effects may be relevant in cosmology and astrophysics
(see, e.g. \cite{CCrunning,Grunning}) and can be explored in
different theoretical frameworks, such as semiclassical gravity
\cite{nelpan82,buch-84}, higher derivative quantum gravity
\cite{frts82,salstr}, low-energy effective quantum gravity
\cite{Don94}, induced gravity \cite{CC-ind} and functional
renormalization group \cite{FRG-QG}. Indeed, the status of the
corresponding types of quantum corrections is different, but there
are also some common points. In particular, in many cases one can
formulate general restrictions on the running of the Newton constant
$G$, which are based on covariance and dimensional arguments
\cite{LRL}. The beta function for the inverse Newton constant
which follows from these condition has the form
\beq
\mu \frac{d}{d\mu}\,\frac{1}{G}
=
\sum\limits_{ij} A_{ij} \,m_i  m_j\,,
\label{runningG}
\eeq
where $m_i$ are masses of the fields or more general parameters
in the action with the dimensions of masses and $A_{ij}$ are given
by series in coupling constants of the theory.

In the perturbative quantum gravity case there may be one more
complication. In the model based on Einstein-Hilbert's gravity there
is no beta function for $G$, and in the four-derivative model this
beta function is dependent on the choice of gauge-fixing condition
\cite{frts82,a,percacci,ohta}. Only a dimensionless combination of $G$
and  $\La_{\rm cc}$ has well-defined running, but this is not sufficient
for the mentioned applications to cosmology and astrophysics.

Recently there was a significant progress in development of
perturbative quantum gravity models which have very different
properties. If the action of the theory includes local covariant
terms that have six or more derivatives, this theory may
possess a few remarkable properties, as listed below.
 %% red
First of all, such a theory  is super-renormalizable \cite{highderi}
(see also earlier works \cite{krasnikov,kuzmin}).
Secondly, there are formal arguments showing that the theory
can  be formulated as unitary, in the case that there are only complex
conjugate massive poles in the tree-level propagator
\cite{LM-Sh,LWQG}. Moreover,
 the theory is unitary at any order in  the perturbative loop
 expansion when the CLOP prescription \cite{CLOP,Wise} is
 implemented or non-equivalently when the theory is defined through
 a non-analytic Wick rotation from Euclidean to Minkowskian
 signature. The third important advantage is that the theories
 with more than
six derivatives may have gauge-fixing and field reparametrization
independent beta functions  for both $G$ and  $\La_{\rm cc}$.
There are also serious indications that the problem of stability
of the theory with complex poles can be resolved within the
Lee-Wick approach. In order to avoid lengthy discussion of this
issue (which has no direct relation to the main subject of the
paper) in the Introduction, we devote to it an Appendix C.

All this means that such a theory satisfies the
 minimal set of consistency conditions and deserves a detailed
 investigation at both classical and quantum levels.  The classical
 aspects of the theory started to be explored recently in
 \cite{mod-New}, where it was shown that the version with
 real simple poles has no singularity in the modified Newtonian
 potential. Quite recently this result was generalized
 for more general cases including multiple and complex poles
 \cite{ABSh1,BG-PLB}.
 Furthermore, in \cite{ABSh2,ABSh1} the detailed analysis
 of light bending in six-derivative models was given. Another
 generalization of simple higher derivative model is nonlocal
 gravity, where we allow for nonlocal functions of differential
 operators \cite{modesto}. Moreover, there exists a class of
 nonlocal theories in which UV behaviour is exactly the same
 like in polynomial higher derivative theories. Therefore, they
 also satisfy the above three points, namely they are quantum
 super-renormalizable unitary models and the analysis of divergences
 and RG running presented here apply to these theories as well.

Until now, the unique example of quantum calculations in the
super-renormalizable quantum gravity was the derivation of the
beta function for the cosmological constant in \cite{highderi}.
Here we start to explore the models further and derive the most
relevant phenomenologically one-loop beta function for the
Newton constant $\,G$.

The work is organized as follows. In Sect.~\ref{II} one can
find a brief general review of the super-renormalizable models
\cite{highderi}, including power counting and gauge-fixing
independence of the beta functions. In Sect.~\ref{III}, we
describe  the one-loop calculations. Some of the relevant
bulky formulas are separated into Appendix A, to provide a
smooth reading of the main text. In Sect.~\ref{IV}, two
important classes of the nonlocal models of quantum gravity
are considered. It turns out that the derivation of one-loop
divergences by taking a limit in the results for a polynomial
models meets serious difficulties, which can be solved only
for a special class of nonlocal theories, which are asymptotically
polynomial in the ultraviolet regime. But still these theories are
super-renormalizable or even finite.
In Sect.~\ref{V5}, the renormalization group for the Newton
and cosmological constants are discussed, within the minimal
subtraction scheme of renormalization. Finally, in Sect. ~\ref{VI}
we draw our conclusions and outline general possibilities for
further work.

%%%%%%%%%%%%%%%%%%%%%%%%%%%%%
%%%%%%%%%%%%%%%%%%%%%%%%%%%%%
%%%%%%%%%%%%%%%%%%%%%%%%%%%%%
\section{Six and higher derivative quantum gravity}
\label{II}

%% \begin{widetext}
One of the simplest models of super-renormalizable
quantum gravity is based on the action
\beq
&& S_{N}
 = \int \!d^4x \sqrt{|g|}\,
\Big\{
\om_{N,R} R \, \square^{N} R
+ \om_{N,C} C \, \square^{N} C
\nonumber
\\
&& % \hspace{2.9cm}
 + \om_{N,\mathrm{GB}}\mathrm{GB}_{N}
 + \om_{N-1,R}R \, \square^{N-1}R
 +\om_{N-1,C}C \, \square^{N-1}C
\nonumber
\\
&&
+\om_{N-1,\mathrm{GB}}\mathrm{GB}_{N-1}
+ \, \dots
%\hspace{2.9cm}
 \nonumber
\\
&&
 + \om_{0,R}R^2
 +\om_{0,C}C^2
+\om_{0,\mathrm{GB}}\mathrm{GB}_{0}
 +\om_{\rm EH}R
 +\om_{\rm cc}
 \Big\}\,.
 \label{gaction}
\eeq
Here $\,R\,$ is the scalar curvature and $\,C\,$ is the Weyl
tensor (with writing of all indices suppressed), e.g.
\beq
C \, \square^n\,C
=
\mathrm{Riem} \,
\square^{n}\mathrm{Riem}
- 2\mathrm{Ric} \, \square^{n}\mathrm{Ric}
+ \frac13\,R\square^{n}R\,.
\label{C2n}
\eeq
In the last formula $\mathrm{Riem}$ and $\mathrm{Ric}$ stand for
the Riemann and Ricci tensors (with writing of all indices suppressed
again), correspondingly. Furthermore, the generalized
Gauss-Bonnet (GB) term is
\beq
\mathrm{GB}_{n}
=
\mathrm{Riem}
\square^{n}\mathrm{Riem}
- 4\mathrm{Ric}\square^{n}\mathrm{Ric}
+R\square^{n}R \, ,
\label{GBn}
\eeq
which is not topological for $\,n\neq 0$. It can be easily shown that
this term can be reduced to the $\,{\cal O}({\rm Riem}^3)\,$ terms
for these values of $n$. We also notice that for $n=0$ and in $d=4$
dimensions any order symmetrized variation of the GB term is a total
derivative \cite{hove,capkim} hence it does not contribute at all.
The coefficient of the Einstein-Hilbert term
and the density of the cosmological constant term are denoted as
$\,\om_{\rm EH}=-1/(16\pi G)\,$ and
$\,\om_{\rm cc} = - \La_{cc}/(8\pi G)\,$, in order to provide
homogeneous notation.  Action \eqref{gaction} is the most
general  one which is at most quadratic in curvatures. One can
extend it preserving power counting by adding extra terms that are
of the higher orders in the curvature tensor \cite{modestoLeslaw}.
As we shall discuss in what follows, such
$\,{\cal O}({\rm Riem}^3)$-terms are not necessary for
super-renormalizability and in this sense they represent the
non-minimal sector of the theory. In what follows we will see
that the Einstein-Hilbert and cosmological terms are relevant,
but for  $N\geqslant2$  they do not affect the divergences.

It proves useful to discuss the dimensions of the parameters
included into the action (\ref{gaction}). Standard considerations
show that the dimensions of the parameters are
\beq
&& \big[\om_{N,R}, \,\, \om_{N,C}, \,\, \om_{N,\mathrm{GB}}\big]
=({\rm mass})^{-2N} \, ,
 \\
&& %\mbox{and}
\big[\om_{N-1,R},\,\, \om_{N-1,C},\,\, \om_{N-1,\mathrm{GB}}\big]
=({\rm mass})^{-2N+2}\,.
\nonumber
\eeq
Therefore, by dimensional reasons similar to the ones that lead
to (\ref{runningG}), one can show that the beta function for the
inverse Newton constant should be proportional to the ratios
$\,\om_{N-1,i}/\om_{N,j}$, where $\,i,j=(R,C,{\rm GB})$
for $N\geqslant2$. In what follows this  conclusion will be
supported by power counting and by the direct calculation of
the one-loop counterterms.

Let us consider the power counting in the theory (\ref{gaction})
which is equivalent to the most general local theory with $\,2N+4\,$
derivatives of the metric. The formula for the superficial degree of
divergence $D$ for a $p$-loop diagram follows from a general
expression
\beq
D + d
=
\sum\limits_{l=1}^{l_{\rm int}}(4-r_l)
-4n_v +4+\sum\limits_{v}K_v
\eeq
and the topological relation
\beq
l_{\rm int} = p + n_v - 1\,.
\label{topo}
\eeq
Here $d$ is the number of derivatives on the external lines of
the diagrams, $r_l$ is the power of momenta in the inverse of
the propagators, $K_v$ is the power of momenta in the
given vertex $v$, $n_v=\sum_v$ is the total number of vertices
and $\,l_{\rm int}\,$ is the number of internal lines of the diagram.
One can introduce the gauge-fixing condition in the theory
(\ref{gaction}) in such a way that effectively $\,r_l \equiv 2N+4$
(for this one can check \cite{highderi} and the next section of the present
work for the details).
The strongest divergences come from the largest $\,K_v\,$ and
hence one can set  $\,K_v=r_l = 2N+4$, without losing generality.
Then it is easy to arrive at the result
\beq
D + d
&=&
4+2N(1-p)\,.
\label{power}
\eeq
For the logarithmic divergences ($D=0$) we get the estimate
for the dimension of the $p$-loop counterterms, $\,d=4+2 N(1-p)$.
The last expressions have several important consequences.

First of all, from  (\ref{power})  one learns that the theories with
$\,N \geqslant1\,$ are super-renormalizable, since only diagrams  with
$\,p=1,2,3\,$ may be divergent. Moreover, for $N \geqslant3$ only
one-loop diagrams may be divergent. As a result, the one-loop
beta functions will be exact in the the theories with $N \geqslant3$.

Second, since the divergences come from $\,p \geqslant1$, it is clear
that for any $\,N \geqslant1\,$ the dimension of the counterterms can
only be $\,d=4,2,0$. As a result, the terms which are included in the
first two lines of the action (\ref{gaction}) are not subjected to the
renormalization procedure for $N\geqslant2$. The multiplicative
renormalizability requires that the action \eqref{gaction} includes
the terms in the last line, which we can also parametrize as
\beq
S_{\rm add}
=
S_{\rm EH}+S_{4D}\,,
\label{vac}
\eeq
where
\beq
S_{\rm EH}
\,=\,
-\frac{1}{16\pi G}\int\!d^4x\sqrt{|g|}\,\big\{R + 2\La_{\rm cc}\big\}
\label{EH}
\eeq
is the Einstein-Hilbert action with the cosmological constant and
\beq
S_{4D}=\!\int\!d^4x\sqrt{|g|}\big\{
a_1C^2+a_2E+a_3{\Box}R+a_4R^2\big\}\,
\label{4D}
\eeq
is the action with four-derivative terms. We included
the surface term $\Box R$, along with the Euler density term $E$.

Third, for $\,N \geqslant2\,$ the  renormalizations of the parameters
$\,a_{1,2,3,4}\,$  and $G,\,\La_{\rm cc}$ of the actions (\ref{EH})
and (\ref{4D}) do not depend on these parameters of the action, but
only on the parameters of the higher derivative terms  of the action
(\ref{gaction}) (first two lines of it). Let us note that the
renormalization of $\La_{\rm cc}/(8\pi G)\,$ was described in
details in \cite{highderi} and therefore we will not discuss its
details here. As far as we intend to calculate the beta function
for the Newton constant $G$, it is sufficient to consider only the
quantum effects of the theory with the action (\ref{gaction}).
Indeed, the result can be modified by adding
$\,{\cal O}({\rm Riem}^3)$-type terms
\cite{highderi,LWQG,modestoLeslaw,universality,CountGhost}
(sometimes called ``killers'' \cite{modestoLeslaw} for  the ability
to make the theory free of any divergence), but we will not consider
these terms here. One can note that the remaining terms, which do
not affect the divergences, can provide the complex massive poles
and hence lead to the unitarity in the Lee-Wick sense. Therefore, the
theory  (\ref{gaction}) turns out to be quite general and deserves an
explicit calculation of the physically relevant beta function for $G$.

Fourth, an important observation is that for $\,N \geqslant1\,$ the
classical equations of motion in the theory have $\,4+2N \geqslant6\,$
derivatives, while the divergences are removed by the counterterms
which have $0$, $2$ and $4$ derivatives. It is well-known that the
gauge-fixing dependence of the one-loop counterterms is always
proportional to the classical equations of motion. The general
 proof of this statement can be found in \cite{VLT82}, and one can
find a specific proof for the one-loop contributions in the book
\cite{book}.
The general theorem enables one to establish gauge dependence
in second- and fourth-order quantum gravity without explicit
calculations \cite{frts82,a}. And in the case of six- or
higher than six-derivative models the same statements guarantee
the gauge-fixing independence of all divergences.

Finally, we conclude that it is completely consistent to calculate
only the beta function for $G$, without deriving the beta functions
for the coefficients of the four-derivative terms $a_{1,2,3,4}$.
Moreover, the calculation of the beta function for $G$ can be
performed in any useful gauge, because the result does not
depend on the choice of gauge-fixing.

%%%%%%%%%%%%%%%%%%%%%%%%%%%%%
%%%%%%%%%%%%%%%%%%%%%%%%%%%%%
%%%%%%%%%%%%%%%%%%%%%%%%%%%%%
\section{One-loop calculation}
\label{III}

The calculation of one-loop divergences can be performed within
the background field method. The splitting into background and
quantum metric is as follows:
\beq
g_{\mu\nu}
&\rightarrow&
g^\prime_{\mu\nu} = g_{\mu\nu} + h_{\mu\nu}\,.
\label{bfm}
\eeq
The Lagrangian quantization is based on the Faddev-Popov
scheme requires introducing the gauge-fixing condition
$\chi_{\mu}$ and the weight operator $C^{\mu\nu}$.

The one-loop effective action is given by expression
\cite{frts82}
\beq
%\hspace{5cm}
\bar{\Ga}^{(1)}(g_{\mu\nu})
&=&
\frac{i}{2}\,\ln \Det \hat{\cal{H}}
-
\frac{i}{2}\,\ln \Det \,\hat{\tilde{C}}
\nonumber
\\
&-&
i\,\ln \Det \hat{\cal{H}}_{\rm gh}\,,
\label{e}
\eeq
where $\,\hat{\cal{H}}\,$ is the bilinear form of the action
(\ref{gaction}) with the gauge-fixing term added with respect
to gravitational perturbations $h_{\mu\nu}$, and
$\hat{\cal{H}}_{\rm gh}\,$ is the bilinear form of the action of
the gauge ghosts. $\hat{\tilde{C}}$ is the bilinear form of the
gauge-fixing action with respect to gauge-fixing conditions,
which is related to the weight operator. All operators in (\ref{e})
are second variational derivatives of the corresponding actions
and therefore are Hermitian.
Moreover, they should be divided by the corresponding powers
of the renormalization parameter $\,\mu\,$ in the dimensional
regularization, which are not shown in the last formula. Also,
in order to have homogeneous propagators for the quantum
metric and the ghosts, one can, e.g. replace
$\,\hat{\cal{H}}_{\rm gh} = M_\al{}^\be\,$
by the product
$\,\tilde{C}^{\ga\al}\,M_\al{}^\be$,
while the coefficient of the weight operator contribution in (\ref{e})
changes from $-i/2$ to $i/2$.

In this paper the one-loop computations have been done in
Euclidean signature.
%%% red
Indeed, for the Lee-Wick gravitational theories, which we proposed
for the first time in \cite{LM-Sh,LWQG}, we can have extra nonlocal
divergences in Minkowski space due to the Wick rotation in the theory
with complex poles. It is important to note that these divergences
have nothing to do with the UV limit and renormalization group which
we deal with here. At the same time, no problems of this kind exist
for the theories with real poles.

Let us expand on the case of complex conjugate poles.
For Lee-Wick theories the continuation
to Minkowski space is guaranteed consistently with perturbative
unitarity by the procedure described below.

The CLOP prescription \cite{CLOP} (due to  Cutkosky, Landshoff,
Olive, and Polkinghorne)  gives a Lorentz-invariant and unitary result.
It consists
on taking the masses of the complex conjugate poles to be unrelated
complex mass parameters to avoid the overlap of the poles for finite
values of the external energy. At the end of the computation one has
to impose the condition that the poles are complex conjugate to
each other. In particular, the CLOP prescription allows to make the
Wick rotation to Euclidean signature, as explained in \cite{Wise}.
The CLOP prescription does not affect the renormalization of the
theory because the divergent contributions to the quantum effective
action have a regular limit when we impose the square masses to be
complex conjugate each other. The CLOP prescription gives an
unambiguous result in simple diagrams, as confirmed by the explicit
calculation of the bubble diagram in \cite{Wise}. However, in
\cite{CLOP} it is shown that there are ambiguities at higher order
in the loop expansion.

Let us note that the described complications concern only the
case when the massive poles are complex conjugate, while for
the local theories with real ghosts there is no need to worry
about the Lee-Wick quantization or about continuation to
Minkowski space, since unitarity is already violated at tree-level.

%%%%%%%%%%%%%%%%%%%%%%%%%%%%%
%%%%%%%%%%%%%%%%%%%%%%%%%%%%%
\subsection{Gauge-fixing and relevant operators}

The general form of the gauge-fixing action is
\beq
S_{{\rm gf}}
=
\int\!d^4x\sqrt{|g|}\,\,
\chi_{\mu}C^{\mu\nu}\chi_{\nu}\,,
\label{gf}
\eeq
where the gauge condition is
\beq
\chi_{\mu}
=
\nabla^{\lambda}h_{\lambda\mu}-\beta\nabla_{\mu}h\,.
\label{chi}
\eeq
Here and below the covariant derivatives are constructed with
the background metric. It is sufficient for us to take the
weight function $C^{\mu\nu}$ in the following form
\beq
C^{\mu\nu}
=
-\,\frac{1}{\alpha}\left(g^{\mu\nu}\square
+\gamma\nabla^{\mu}\nabla^{\nu}
-\nabla^{\nu}\nabla^{\mu}\right)\square^{N}\,.
\label{weight}
\eeq
By $\tilde C^{\mu\nu}$ in \eqref{e} we mean the self-adjoint
extension of the operator $C^{\mu\nu}$. The action of the gauge
ghosts $\bar{c}^\mu$ and $c_\nu$ has a standard form
\beq
S_{{\rm gh}}
=
\int\!d^4x\sqrt{|g|}\,\,{\bar c}^{\mu}\,M_\mu{}^\nu\,c_{\nu}\,,
\label{ghosts}
\eeq
the operator $M$ is defined as usual (see, e.g., \cite{book}, because
in this part there is no essential difference with the four-derivative
quantum gravity), such that we arrive at
\beq
M_{\mu}{}^{\nu}=\delta_{\mu}^{\nu}\square+\nabla^{\nu}\nabla_{\mu}-2\beta\nabla_{\mu}\nabla^{\nu}\,.
\label{ghop}
\eeq
At that point one can observe the first essential simplification
coming from our interest in the Einstein-Hilbert counterterm only.
For the gauge-fixing action (\ref{gf}) with (\ref{chi}) and
(\ref{weight}), both the weight operator $\tilde{C}^{\mu\nu}$ and
the bilinear operator of the ghost action $M_\mu{}^\nu$ in
(\ref{ghop}) are homogeneous functions of covariant derivative
$\na_\al$ and do not include dimensional parameters. As a result,
the corresponding
functional determinants contribute only to the divergences of the
four-derivative terms in (\ref{4D}). Since we are interested in the
beta function for the Newton constant only, we can concentrate our
attention only on the first term in the expression (\ref{e}) and do not
pay attention on the other two terms.

Let us consider the relevant first term in (\ref{e}).  We remember, once
again, that the final result for divergences does not  depend on the
choice of $\chi_{\mu}$ and $C^{\mu\nu}$. Therefore, these functions
can be defined to make calculations as simple as possible. The
primary role of the gauge-fixing is to make the operator
$\hat{\cal H}$ non-degenerate and this can be achieved for different
values of the gauge-fixing parameters $\al$, $\be$, and $\ga$ in
Eqs.~(\ref{chi}) and (\ref{weight}). Due to the gauge-fixing
independence of the one-loop effective action one can choose the
gauge-fixing parameters in such a way that operator $\hat{\cal H}$
assumes the  minimal form. This means that the highest-order
derivatives form the combination $\,\square^{N+2}$. The
corresponding calculations are essentially the same as in the
four-derivative quantum gravity.

The requirement of minimality leads to the following values:
\beq
\al
&=&
\frac{2}{\omega_{N,C}}\, ,
\quad
\beta =
\frac{\om_{N,C} - 6\omega_{N,R}}{4\omega_{N,C}
- 6\omega_{N,R}}\,,
\nonumber
\\
\ga
&=&
\frac{2\omega_{N,C}-3\omega_{N,R}}{3\omega_{N,C}}\,.
\label{mini}
\eeq
One can see that the gauge-fixing parameters $\be$ and $\ga$ depend
only on the ratio $\,\om_{N,R}/\om_{N,C}\,$ and are independent on
other coefficients appearing in the gravitational action
(\ref{gaction}). In particular, the minimal choice of the gauge-fixing
conditions does not depend on the generalized Gauss-Bonnet term,
namely $\,\om_{N,\mathrm{GB}}$. This is  obviously a natural output
because the minimal gauge-fixing can be defined on a flat background,
where the $\,{\cal O}({\rm Riem}^3)\,$ terms are irrelevant.
Furthermore, the parameter $\al$ is only dependent on $\,\om_{N,C}$.
This is because only the term $\,C\square^{N}C\,$ in the action is
responsible for the propagation of the spin-2 part of gravitational
perturbations on flat spacetime with the highest number of derivatives.

The evaluation of the main expression
\beq
\frac{i}{2}\,\ln \Det \hat{\cal{H}}
=
\frac{i}{2}\Tr \ln H^{\mu\nu,\,\rho\sigma}
\eeq
can be performed by using the generalized Schwinger-DeWitt technique
developed by Barvinsky and Vilkovisky in \cite{barvi85}, in a way
qualitatively similar to the calculation of the cosmological constant
counterterm in \cite{highderi}. Technically, the calculation of the
 linear in $R$ term is much more complicated, and we shall describe
 it in certain details. The minimal choice of the gauge parameters
 (\ref{mini}) provides the useful form of the operator \cite{highderi}
\beq
H^{\mu\nu,\,\rho\sigma}
=
{\cal B}^{\mu\nu,\,\ka\la}\,\,H'_{\ka\la}{}^{\rho\si}\,,
\label{H}
\eeq
where $\,{\cal B}^{\mu\nu,\,\ka\la}\,$ is the DeWitt metric in
the space of the fields for our model,
\beq
{\cal B}^{\mu\nu,\,\ka\la}
= \frac{\om_{N,C}}{2}\,\Big[
\de^{\mu\nu,\,\ka\la}
- \frac{\om_{N,C}
- 6\om_{N,R}}{2\big(2\om_{N,C}-3\om_{N,R}\big)}
\,g^{\mu\nu}\,g^{\ka\la}\Big],
\nonumber
\\
\eeq
where
\beq
\de^{\mu\nu,\,\ka\la}
=
\frac12\,\big( g^{\mu\ka} g^{\nu\la} + g^{\mu\la}g^{\nu\ka}  \big)
\nonumber
\eeq
is the identity matrix in the space of the fields. In what follows we
assume that the $c$-number operator ${\cal B}^{\mu\nu,\,\ka\la}$
is non-degenerate, such that its inverse
\beq
{\cal B}_{\al\be,\,\mu\nu}^{-1}
=
\frac{2}{\om_{N,C}}\,\left[
\de_{\al\be,\,\mu\nu}
+
\frac{\om_{N,C} - 6\om_{N,R}}{18\,\om_{N,R}}
\,g_{\al\be}\,\,g_{\mu\nu} \right]
\label{inverseB}
\eeq
has finite coefficients. Let us note that the degenerate case, when
$\,\om_{N,C}=0\,$ or $\,\om_{N,R}=0\,$, is non-renormalizable,
and therefore is much less interesting.

Acting on (\ref{H}) with the inverse operator (\ref{inverseB}),
one can see that the operator $\,H'_{\ka\la}{}^{\rho\si}\,$ has
the standard minimal form with the senior term of exactly $2N+4$
derivatives, and with the identity matrix in the space of the fields
\beq
\de^{\rho\si}_{\ka\la}=\frac{1}{2}(\de^\rho_\ka\de^\si_\la
+\de^\rho_\la\de^\si_\ka)
\nonumber
\eeq
in front instead of the  DeWitt
metric $\,{\cal B}^{\mu\nu,\,\ka\la}\,$ as in \eqref{H}:
\beq
H'_{\ka\la}{}^{\rho\si}
&=& %\hspace{-1.2cm}
\delta^{\rho\si}_{\ka\la}\,\square^{N+2}
\nonumber
\\
&+&
V_{\ka\la}{}^{\rho\si,\,\la_1\ldots\la_{2N+2}}
\na_{\la_1}\cdots\nabla_{\la_{2N+2}}
\nonumber
\\
&+&
W_{\ka\la}{}^{\rho\si,\,\la_1 \ldots\la_{2N+1}}
\na_{\la_1} \cdots \na_{\la_{2N+1}}
\\
&+&
U_{\ka\la}{}^{\rho\si,\,\la_1 \ldots \la_{2N}}
\na_{\la_1} \cdots \na_{\la_{2N}}+
{\cal O} \big(\na^{2N-1}\big)\,.
\label{Hprime}
\nonumber
\eeq

Obviously, the first factor in Eq. (\ref{H}) does not contribute to
the divergent part of the effective action, and therefore for us
\beq
\frac{i}{2}\,\Tr \ln H^{\mu\nu,\,\rho\sigma}
%% \right|_{{\rm div}}
=
%% \left.
\frac{i}{2}\,{\rm Tr}\, \ln\, H'_{\ka\la}{}^{\rho\si}\,.
%% \right|_{{\rm div}}
\eeq
Different from the calculation of the cosmological constant
counterterm \cite{highderi}, the expressions for $U$ and $V$
requested for the  derivation of divergences for an arbitrary metric
background are very bulky and difficult to derive and to deal with.
However, there are possibilities of great simplifications when we
are interested only in the counterterm linear in scalar curvature.
First of all, by dimensional reasons the mass dimensions of $V,W$
and $U$ are $2,3$ and $4$, correspondingly.
 If only coefficients $\om_{N,i}$ ($i=R,C,{\rm GB}$) entered
 the expression for $H'$ tensor, then only $U$ could be relevant
 in our case. However, the other coefficients $\om_{N-1,i}$ enter
 too, hence both $V$ and $U$ parts are relevant here.
Furthermore, since we do
not pretend to calculate the $\square R$-term, in the course of
calculations we can assume that the scalar curvature $R$ is a
constant, without losing the generality. Finally, it is good to
remember that the Ricci and Riemann tensors enter the final
expressions for the  linear in scalar curvature term only through
their contractions with the metric tensor. Therefore, again without
loss of generality we can assume that the curvature tensor is
\beq
R_{\mu\nu\al\be} =
\frac13\,\La\,
\big( g_{\mu\al}g_{\nu\be} - g_{\nu\al}g_{\mu\be} \big)\,,
\label{con curv}
\eeq
where $\La$ is a constant. This is equivalent of assuming that the
background is a maximally symmetric spacetime (dS or AdS) with
the radius of curvature given by $\pm\sqrt{3/\La}$. Due to the
covariant constancy of the purely background metric we have $W=0$.
%and we shall assume that this is the case.
At the end one can set  $\,\La=R/4$ and arrive at the general result
of the counterterm in the one-loop effective action, which is valid for
both constant and non-constant scalar curvature $R$, in the given
approximation. On the basis of dimensional analysis, in the tensor
$\,V\,$ there are terms of zero and first powers in the background
curvature $\La$, that can be written schematically as
\beq
V \,=\, {\tilde V}_{(0)}
+ \La {\tilde V}_{(1)}\,,
\label{V}
\eeq
while for the tensor $U$ there are terms of the following types
\beq
U =
{\tilde U}_{(0)}
+ \La {\tilde U}_{(1)}
+ \La^2 {\tilde U}_{(2)}
\label{U}
\eeq
that means up to the second power in background curvature.
The dimensions of the different types of terms are compensated by
the coefficients, which may have different powers of dimensional
parameters of the action (\ref{gaction}), e.g. by the ratios
$\om_{N-1,i}/\om_{N,j}$, where $i,j=R,C,$ or ${\rm GB}$.
The details of this structure can be found in the Appendix A.

In order to use the generalized Schwinger-DeWitt technique of
\cite{barvi85}, we can first extract the highest derivatives in the
form of $\square^{N+2}$ \cite{highderi},
\beq
\Tr \ln H'_{\ka\la}{}^{\rho\si}
&=& \Tr \ln \square^{N+2} +
\Tr \ln
\Big\{ \de_{\ka\la}^{\rho\si}
\nonumber
\\
&+&
V_{\ka\la}{}^{\rho\si,\,\al_{1}\ldots\al_{2N+2}}
\na_{\al_1}\cdots\na_{\al_{2N+2}}\,\frac{1}{\square^{N+2}}
\nonumber
\\
&+&
W_{\ka\la}{}^{\rho\si,\,\al_{1}\ldots\al_{2N+1}}\,
\na_{\al_1}\cdots\na_{\al_{2N+1}}\frac{1}{\square^{N+2}}
\nonumber
\\
&+&
U_{\ka\la}{}^{\rho\si,\,\al_1 \ldots \al_{2N}}
\na_{\al_1} \cdots \na_{\al_{2N}}\,\frac{1}{\square^{N+2}}
\nonumber
\\
&+&
{\cal O} \Big(\na^{2N-1}\,\,\frac{1}{\square^{N+2}}\Big)\Big\}\,
\eeq
and after that perform a series expansion of the logarithm. The first
term in the last expression does not produce linear in  curvature
divergences and the last term includes terms that cannot give
relevant contributions by dimensional reasons. In what follows we
omit the last term. The expression can be further elaborated as
follows, reducing to the universal traces of \cite{barvi85},
\beq
&& %\hspace{-1cm}
\Tr \ln H'_{\ka\la}{}^{\rho\si}
= (N+2) \Tr \ln\square
\nonumber
\\
&&
+ \Tr \Big\{ V_{\ka\la}{}^{\rho\si,\,\al_{1}
\ldots\al_{2N+2}}\na_{\al_{1}}\dots\na_{\al_{2N+2}}
\,\frac{1}{\square^{N+2}}
\Big\}
\nonumber
\\
&& %\hspace{1cm}
+
\Tr
\Big\{
W_{\ka\la}{}^{\rho\si,\,\al_{1}\ldots\al_{2N+1}}
\na_{\al_{1}}\dots\na_{\al_{2N+1}}\,\frac{1}{\square^{N+2}}
\Big\}
\nonumber
\\
&&%\hspace{1cm}
+ \Tr \Big\{
U_{\ka\la}{}^{\rho\si,\,\al_{1}\ldots\al_{2N}}
\na_{\al_{1}}\dots\na_{\al_{2N}}\,\frac{1}{\square^{N+2}}
\Big\}
\nonumber
\\
&& %\hspace{1cm}
- \frac{1}{2}\Tr
\Big\{ V_{\ka\la}{}^{\mu\nu,\,\al_1\ldots\al_{2N+2}}
V_{\mu\nu}{}^{\rho\si,\,\be_{1}\ldots\be_{2N+2}}
\nonumber
\\
&&
\times \na_{\al_{1}}\dots\na_{\al_{2N+2}}\na_{\be_{1}}
\dots\na_{\be_{2N+2}}\,\frac{1}{\square^{2N+4}}\Big\}  .
\label{expa}
\eeq
Let us start the analysis of the last expression with two simple
observations. The commutators of covariant derivatives with $V$
and $\square$ to negative power operators
in the last term were disregarded because they are irrelevant in the
linear in curvature approximation, as it was explained above.
According to the Eq. (\ref{expa}) and the table of universal
traces of \cite{barvi85}, the trace of the operator proportional to
the tensor $W$ vanishes identically, which confirms our previous
consideration. Therefore, we need to consider only linear in
$V$ and $U$ single traces and the quadratic in $V$ mixed
(interference) trace. Due to the fact that in the tensors $V$ and $U$
we  have at most four free covariant derivatives (not contracted with
other  derivatives) we can simplify the relevant parts of the
expression  \eqref{expa} to the form
\beq
%\hspace{-2cm}
\Tr \ln H'_{\ka\la}{}^{\rho\si}
&=&
\Tr
\Big\{
V_{\ka\la}{}^{\rho\si,\,\al\be\ga\de}\na_\al\na_\be\na_\ga\na_\de
\,\frac{1}{\square^3}
\Big\}
\nonumber
\\
&+&
\Tr \Big\{
U_{\ka\la}{}^{\rho\si,\,\al\be\ga\de}\na_\al\na_\be\na_\ga\na_\de
\,\frac{1}{\square^4}
\Big\}
\nonumber
\\
&-&
\frac{1}{2}\Tr \Big\{
V_{\ka\la}{}^{\mu\nu,\,\al_1\be_1\ga_1\de_1}
V_{\mu\nu}{}^{\rho\si,\,\al_2\be_2\ga_2\de_2}
\nonumber
\\
&\times &
\na_{\al_1}\na_{\be_1}\na_{\ga_1}
\na_{\de_1}\na_{\al_2}\na_{\be_2}\na_{\ga_2}\na_{\de_2}
\,\frac{1}{\square^6}\Big\}
\label{expashort}
\nonumber
\\
&&
+\,\dots\,,
\eeq
where we omitted terms which do not contribute  to  the
divergences  linear  in $R$.
The explicit expressions for the $c$-number tensors $U$ and $V$
can be found in Appendix A.

We here make some comments on the correct way of taking the
traces that are in general not an easy task. It is important to note that
the expression above includes all the divergences that are linear in
$R$. A special care must be given to the interference term
$-\frac{1}{2}{\rm Tr}V^{2}$. Due to dimensional reasons we
are interested in the mixing terms only between the parts of the
$V$ tensor that are linear in $\La$ or independent on $\La$.
Only these terms can give the correct dimension of the coefficient
of $R$ as well as the correct power
of background curvature. Then for the mixing terms we do not
need to account for additional background curvature arising
from commutations. Furthermore, when taking the single trace of $V$
we need only terms independent on $\La$. Contrary, in the traces that
involve the tensor $U$ we are interested
only in those parts that are proportional to $\La$.

The final technical comments are the following.
The interesting parts, which give divergent contributions linear in
curvature, of $V$ and $U$ are linearly proportional to the couplings
$\om_{N-1,C}$, $\om_{N-1,R}$
or $\om_{N-1,{\rm GB}}$ and originate from the variation of the
corresponding part of the  action (\ref{gaction}). These
contributions must be independent on $\La$ in the $V$ part
and linear in $\La$ in the $U$ part, while the denominators with
$\om_{N,R}$ or $\om_{N,C}$ emerge only because of the inverse
DeWitt metric (\ref{inverseB}). To derive the relevant $U$ term
correctly one has to take into account the commutations up to the
order linear in background curvature in the second variation of the
gravitational action \eqref{gaction}.

The situation with the interference term is different. The first
interference factor in $V$ has the same origin as in the single trace
discussed above. However, the second one must be precisely
linear in $\La$ and independent on the couplings $\om_{N-1,i}$.
It contains some rational functions of the dimensionless ratios
between the coefficients $\om_{N,R}$, $\om_{N,C}$ and
$\om_{N,{\rm GB}}$. The origin of this term is either from the
second variation of the gravitational action or from the gauge-fixing
term, both multiplied by the inverse DeWitt metric. To derive this
term correctly one has to take into account the commutations up
to the order linear in background curvature both in the second
variation of the gravitational action or the gauge-fixing term.
In the intermediate expressions one can observe rather complicated
rational functions, and only in the final traces there are
essential simplifications. The precise expressions for relevant
parts of tensors $V$ and $U$ are not given here due to their
lengths.

Actually in \eqref{expashort} we will never need to take universal
trace with more than four free covariant derivatives because of the
structure of the indices and because only the metric functions appear
in the $V$ and $U$ tensors. Therefore, the last but one term in
\eqref{expashort} will never be so long after explicit contractions.

%%%%%%%%%%%%%%%%%%%%%%%%%%%%%
\subsection{Divergences for $N\geqslant2$ models}

After taking the traces according to \cite{barvi85}, we can write
the results for the divergent part of the one-loop effective action. The
 part linear in curvature is given by the expression
\beq
&&
 \Ga^{(1)}_{\rm EH,\,div}(N,\,{\rm GB})
\, = \,\frac{\mu^{n-4}}{  2 (4 \pi)^2 } \frac{1}{\vp}
\int\!  d^nx\, \sqrt{|g|}\,
\nonumber
\\
&&
\times
%% (-1)
\bigg\{ \frac{5 \om_{N-1,C}}{6\om_{N,C}}
+ \frac{\om_{N-1,R}}{6\om_{N,R}}
\\
&&
+(2N-1)\left[
\frac{5 \,\om_{N-1,C}}{6\om^2_{N,C}}
-\frac{\om_{N-1,R}\,}{18\om^2_{N,R}}
\right]
\,\om_{N,\mathrm{GB}}
\nonumber
\\
&&
+ (2N-3)\left[\frac{5}{6\om_{N,C}}
- \frac{1}{18\om_{N,R}}
\right]\,\om_{N-1,\mathrm{GB}}
\bigg\} R,
\nonumber
\label{divsGB}
\eeq
where $\vp=(4-n)/2$  is the parameter of dimensional
regularization, $\mu$ is the dimensionful regularization parameter
and  by $n$ we denote spacetime dimensionality. One can
see that the generalized Gauss-Bonnet terms with the coefficients
$\om_{N,\mathrm{GB}}$ and $\om_{N-1,\mathrm{GB}}$ give
non-zero contributions in the formulas above, differently from the
four-derivative quantum gravity case \cite{Gauss,Weyl}.

Let us stress that the result for the divergence (\ref{divsGB})
is valid only for $N\geqslant2$ and there is no smooth limit $N\to 0$
or even $N\to 1$.
The reason for this discontinuity is that the expansion of
$\square$ into the quantum field $h_{\mu\nu}$ generates
expressions of the $\square^{-1}$ and  $\square^{-2}$ type,
which are badly defined when acting on a constant. As a result
it becomes impossible to take the limits $N\to 0$ and $N\to 1$
in the bilinear form of the generalized Gauss-Bonnet term.
After this expansion the formulas do not ``recognize'' that for  $N=0$
this term is topological. This is why the coefficients of the
$\om_{N-1,\mathrm{GB}}$  and $\om_{N,\mathrm{GB}}$
terms are proportional to $2N-1$ and $2N-3$, and not to $N$,
as one might have expected from them. Indeed, the same discontinuity
takes place also for the cosmological constant - type divergence
derived in \cite{highderi}. We have checked independently
that for $N=0$ the Gauss-Bonnet term does not contribute to
the divergences. The details are not shown here, because one can
find them in the previous works \cite{Weyl} and \cite{Gauss}.
We also remark that even for the calculations in highly nonlocal theory
with the expressions of the $\square^{-1}$ and  $\square^{-2}$ type
for one-loop divergences one is not required to impose proper
boundary conditions for these operators. This detail has been known
from long time ago, e.g., from the pioneer paper by Ichinose
\cite{Ichinose}.

If the relevant generalized Gauss-Bonnet terms are absent,
$\om_{N-1,\mathrm{GB}}=0$  and $\om_{N,\mathrm{GB}}=0$,
we arrive at a very compact form of divergences,  which is valid for
$\,N\geqslant 2$,
 \beq
&&
\Ga^{(1)}_{\rm EH,\,div}
\,= \,\frac{\mu^{n-4}}{12 (4 \pi)^2 \vp}
\int \! d^nx
\sqrt{|g|}\,
%% \left(- \frac16 \right)
\Big\{
\frac{5\om_{N-1,C}}{\om_{N,C}}
\nonumber
\\
&&
+\frac{\om_{N-1,R}}{\om_{N,R}}\Big\}\,R
%% \nonumber \\ &&
%  \hspace{1.25cm}
\,\equiv\,
-\,\frac{\mu^{n-4}}{2\vp}
\!\int\!  d^4x\, \sqrt{|g|}\, \beta_{G} R
\,.
\label{EHdiv-simp}
\eeq
One can note that in the last expression the number of derivatives
$N$ does not appear explicitly, but only through the coefficients of
the action (\ref{gaction}). This is in contrast to the more general
formula \eqref{divsGB}, where the explicit dependence on $N$
takes place in the sector related to the generalized Gauss-Bonnet
terms.

For the sake of completeness, let us write down the expression
for the cosmological constant  divergences, derived in
\cite{highderi}, in the notations which we use here,
\beq
&& \Ga^{(1)}_{\rm cc,\,div}
= - \frac{\mu^{n-4}}{2 (4 \pi)^2} \frac{1}{\vp}  \int \!d^n x
\sqrt{|g|}
\Big(
\frac{5 \omega _{N-2,C}}{\omega _{N,C}}
\nonumber
\\
&&
+\frac{\omega _{N-2,R}}{\omega _{N,R}}
-\frac{5 \omega _{N-1,C}^2}{2 \omega _{N,C}^2}
-\frac{\omega _{N-1,R}^2}{2 \omega_{N,R}^2}
\Big)
\nonumber \\
&&
% \hspace{1.13cm}
    \equiv -
\frac{\mu^{n-4}}{2 \vp} \! \int \!d^4 x
\sqrt{|g|}\, \beta_{\Lambda_{\rm cc}}
   \, .
\label{CCdivs}
\eeq
The contribution of the generalized GB term to the cosmological
constant divergence is zero because this term does not contribute
to the propagator around flat spacetime. It is useful to remember
that both expressions  (\ref{EHdiv-simp}) and (\ref{CCdivs}) are
valid only for $\,N \geqslant2$. It is easy to see that the relative
factor of ``5" between the contributions of tensor and scalar modes
in Eq. (\ref{EHdiv-simp}) is exactly the same which one could
observe in the cosmological constant divergence (\ref{CCdivs}).

It is clear from Eq. ~(\ref{EHdiv-simp}) that one can provide zero
divergence in the Einstein-Hilbert sector by adjusting the
highest-derivative coefficients of the classical action
$\,\om_{N-1,C}$, $\,\om_{N,C}$, $\,\om_{N-1,R}\,$ and
$\,\om_{N,R}$. This can be achieved even without using the
freedom to choose the coefficients of the
$\om_{N-1,\mathrm{GB}}$  and $\om_{N,\mathrm{GB}}$
in (\ref{divsGB}). Moreover, one can make the same for both the
cosmological and the linear in $R$ divergences at the same time.

One can make zero the two remaining relevant coefficients for the
$C^2$ and $R^2$ counterterms by means  of the ``killer'' operators.
Thus one can achieve the one-loop finiteness, which can be directly
extended into all-loop order for the case of $\,N \geqslant3$, since
in this case divergences exist only at the one-loop level.

%%%%%%%%%%%%%%%%%%%%%%%%%%%%%
\subsection{The special cases of $N=1$ and $N=0$}

In order to achieve a better understanding of the situation with
smaller $N$, let us present the corresponding results, without
going into full details. As we have explained above, the cases of
$N=0$ and $N=1$ should be considered separately. The
calculations are pretty much the same as for $N \geqslant2$,
therefore, we present only the final results.

For $N=1$ the action includes the following relevant  terms:
\beq
&&
S_{N=1}
= \int d^{4}x\sqrt{|g|}\,\big\{
\om_{1,R}\, R \, \square R
+ \om_{1,C}C \, \square C
\nonumber
\\
&&
+ \om_{1,\mathrm{GB}}\mathrm{GB}_{1}
+\omega_{0,R}R^{2}
+
\omega_{0,C}C^{2}\big\}\,.
\label{N1}
\eeq
The term linear in curvature in the divergent part of the one-loop
effective action is:
\beq
&&
%\hspace{-1.1cm}
\Ga_{\rm div}^{(1)}
=  -
\frac{\mu^{n-4}}{2 (4 \pi)^2} \frac{1}{ \vp}\!\int\! d^nx
\sqrt{|g|}(-1)\Big\{
\frac{5\om_{0,C}}{6\om_{1,C}}
+ \frac{\om_{0,R}}{2\om_{1,R}}
\nonumber
\\
&&
- \frac{5\om_{0,R}}{\om_{1,C}}
+ \left(
\frac{5\om_{0,C}}{6\om_{1,C}^2}
- \frac{\om_{0,R}}{18\om_{1,R}^2}
\right) \om_{1,\mathrm{GB}}
\Big\} \, R.
\label{N1div}
\eeq
One can easily note that this expression differs from the $N \geqslant2$
case. In particular, there is a mixing of the coefficients for the Weyl and
scalar curvature. Moreover, we notice that only the terms with the Weyl
curvatures and with the generalized Gauss-Bonnet term give rise to
divergences that can be obtained continuously from formula
\eqref{divsGB} in the limit $N=1$.

For $N=0$ the action reduces to the fourth-order quantum
gravity, which was the subject of similar considerations in
\cite{JulTon,frts82,avbar86,Gauss}, namely
\beq
S_{N=0} =
\int d^4x\sqrt{|g|}
\,\left\{
\om_{R}R^{2}+\om_{C}C^2
+\om_{\mathrm{EH}}R \right\}\,.
\label{N0action}
\eeq
We performed the corresponding one-loop calculation just to make
an extra check. The linear in curvature divergent part of the one-loop
effective action is given by
\beq
\Ga_{\rm div}
&=& -\, \frac{\mu^{n-4}}{2 (4 \pi)^2 \,\vp}\,\int  d^nx
 \sqrt{|g|}\,\Big\{
\frac{1}{12\om_{R}}
\nonumber
\\
&-&
\frac{13}{12\,\om_{C}}
\,-\, \frac{5\om_R}{2\om_C^2}\Big\}
\,\om_{\mathrm{EH}}\,R\,.
\label{N0div}
\eeq
Taking into account the difference in notations, this expression
perfectly agrees with the well-known result of \cite{avbar86}.
Let us remember, once again, that this expression depends on
the gauge-fixing \cite{frts82}. The expression (\ref{N0div})
corresponds to the minimal gauge that is the simplest one for
the sake of practical calculations.

Let us present more detailed explanation for the above mentioned
discontinuity between $N=0$, $N=1$ and $N \geqslant 2$ results.
It is worth noticing that there exists another explanation which is
related to the action of the $\square^{-1}$ operator
\cite{Barvinsky:2003kg}. We leave it to the interested reader to
consult this work, and we present here the explanation which is purely
technical.
\
The qualitative reason for the discontinuity is as follows. In the
course of computations of the second variation of the action on a
general background  one has to take into account variations of the
$\square$ operator and its powers. And here is the source of the
discontinuity. When $N=0$, (that means no $\square$  operators)
there is no variation of these operators. When $N=1$ there are
variations of only one $\square$ operator. There may be first or
second variations, but always of one $\square$. Starting from $N=2$
there are qualitatively new terms with two first variations of two
different $\square$ operators appearing in the factor expansion of
the power $\square^N$. We also have to take into account that these
terms, which matter for computation of divergences proportional
to the Ricci scalar $R$, are from level $N-1$, while the exponent
on the box operator leading in UV is $N$, according to
\eqref{gaction}.

These structural differences in taking variations are the reason for
discontinuities
between $N=0$,  $N=1$, and $N=2$ cases, which are present in the
vertices of the theory, or in the expressions for $V$, $W$ and $U$
tensors. However, these differences do not occur in the leading terms in
\eqref{Hprime},
which correspond to the propagator on a flat spacetime background.

Let us note that the structure of vertices explained above is well
presented  in the thesis \cite{eran} for the perturbative calculus
around flat gauge backgrounds and using Feynman technique.
The theory presented there is with action quadratic in field
strengths and containing a form factor of the covariant $\square$
operators.  As we know from \cite{highderi} for the beta function of
the cosmological constant we are interested only in the propagator on
flat spacetime, so there the variations of the $\square$  are not
producing the effect of discontinuity. This effect shows up in the
divergences of the $R$-type which we derived here, and is expected in
the four-derivative divergences as well which are yet to be calculated.
The reason for discontinuities for the beta function of the
cosmological constant case is due to structural changes of the
action. Namely the Einstein-Hilbert term $R$ is neither of the form
$R \square^N R$ nor $C \square^N C$ for none $N$, neither for
$N$ analytically continued to a  domain of negative integers.

%%%%%%%%%%%%%%%%%%%%%%%%%%%%%
%%%%%%%%%%%%%%%%%%%%%%%%%%%%%
%%%%%%%%%%%%%%%%%%%%%%%%%%%%%
\section{Quantum nonlocal models}
\label{IV}

It is tempting to use the results for the divergences in the
polynomial theory (\ref{divsGB}), (\ref{EHdiv-simp}),
(\ref{CCdivs}) in order to obtain the same type of divergences
in a wide class of the nonlocal models of quantum gravity.
The most interesting  theories are certainly the ones that have
no massive ghost-like poles in the propagators. These models
were introduced by Tseytlin \cite{Tseytlin-95} in the framework
of string theory as an alternative to the Zwiebach transformation
of metric \cite{zwie} in the low-energy effective action.
Another interesting feature of the original nonlocal model
\cite{Tseytlin-95} was the absence of singularity in the
Newtonian solution for the gravitational field of a point-like
massive particle. A ghost-free nonlocal gravitational action
was proposed for the first time by Krasnikov in 1988
\cite{krasnikov} and studied by Kuz'min in 1989 \cite{kuzmin}.
Later on, qualitatively similar ghost-free nonlocal models were
suggested by Tomboulis in \cite{Tomboulis-97} as candidates
to be unitary and super-renormalizable or even finite quantum
gravity theories  \cite{modesto,modestoLeslaw,universality}.
One can remember that the general aspects of quantization of nonlocal
quantum field theories have a long history \cite{Efimov}. For
the recent developments in many aspects of nonlocal theories
the reader can follow the literature in
\cite{modesto,modestoLeslaw,universality,scattering,fingauge,exactsol,entanglement,staro,confgr,spcompl,evaporation} and
\cite{Cnl1} about the localization of nonlocal
theories and nonlocal gravity as a  diffusion equation.
Recently in the paper of one of the present authors
\cite{CountGhost} it was shown that an infinite amount
of the ghost-like complex states in these models emerge in
the quantum theory, when one takes the loop corrections into
account. However, perturbative unitarity of the $S$-matrix is
guaranteed by the Cutkosky rules that are not modified by
the kind of non locality that we are considering here.
Nevertheless, it is important to have a better understanding of
the possible form of quantum corrections in such theories.
Let us start the discussion of this problem from the simplest
example \cite{Tseytlin-95}.

Our starting point will be the nonlocal theory of quantum
gravity with the classical action
\cite{krasnikov,kuzmin,Tomboulis-97,modesto}
\beq
S_{\rm nl}
&=&
\int d^4 x \sqrt{|g|} \,
\Big\{ - \frac{1}{\ka^2} \,R
\nonumber
\\
&+&
\frac12\,C\,\Phi\Big(\frac{\Box}{M^2}\Big)C
\,-\,  \frac16 \,R \, \Psi\Big(\frac{\Box}{M^2}\Big) \, R \Big\}\,.
\label{act Phi}
\eeq
We do not add a possible nonlocal generalization of the Gauss-Bonnet
term, but this is possible. The power counting in this theory reduces
to the topological relation (\ref{topo}) between the number of loops,
internal lines and vertices \cite{CountGhost}. Its evaluation shows
the result may be identical to the one of the polynomial theory
(\ref{gaction}) with $N \geqslant3$. This means that the divergences
emerge only at the one-loop order and have the form of the
cosmological constant and Einstein-Hilbert terms,
plus two relevant terms with four derivatives, namely $\,C^2\,$
and  $\,R^2$. In order to achieve this, one has to choose
$\,\Phi(z)\,$ and $\,\Psi(z)\,$ functions to have equal asymptotic
behaviour and be sufficiently fast growing functions at
$\,z \to \infty$. It is clear that we have to deal only with the
model (\ref{gaction}) with a very large $\,N$, hence the result for
the cosmological and linear in $R$ divergences in the polynomial
theory are given by the expressions (\ref{CCdivs}) and
(\ref{EHdiv-simp}).

The most natural option is to choose $\,\Psi\,$ and $\,\Phi\,$ such
that the structures of the Euclidean propagator in spin-2 and spin-0
sectors are the same. Consider the simplest version of the theory \cite{Mtheory,BiswasMazumdar,BiswasKoshelev},
\beq
\Phi = \Psi
= - \,\frac{1}{\ka^2 \, \Box}\big(e^{ - \square/M^2} - 1\big) \,.
\label{PhiPsi}
\eeq
Then we arrive at the equation for the pole
\beq
1 + \ka^2 p^2  \Phi\Big(-\frac{p^2}{M^2}\Big)
=
e^{p^2/M^2}\,,
\label{exp}
\eeq
such that the unique spin-2 pole of the propagator is massless.
In what follows we consider only this version of the theory, but
it is certainly possible to make generalizations.

Let us show, by taking a limit $\,N \to \infty\,$ in the results
(\ref{divsGB}), (\ref{EHdiv-simp}) and (\ref{CCdivs}), that
the coefficients of the one-loop divergences in the model
(\ref{act Phi}) with functions (\ref{exp}) experience an
unrestricted growth. In the polynomial theory the equation
for the poles of the propagator is defined by
\beq
1 + \ka^2 p^2 \,\Phi\Big(-\frac{p^2}{M^2}\Big)
=
\sum\limits_{k=1}^{N}\,\frac{\big(p^2/M^2\big)^k}{k!}
\label{expN}
\eeq
and the same for the function$\,\Psi$. Then the relevant
coefficients of the polynomial action (\ref{gaction}) have
the structure
\beq
\big(\om_{k,C}\,;\,\,\om_{k,R}\big)
\sim \frac{M^{-2k}}{k!},
\quad
k =N,N-1,N-2.
\qquad
\label{omegaNlim}
\eeq

The inspection of the coefficients of expressions (\ref{divsGB})
and (\ref{CCdivs}) shows that they behave like $\,N\,$ and
$\,N^2$, correspondingly. In the limit $\,N \to\infty\,$ both
coefficients explode, that demonstrates a discontinuity of
quantum corrections for the exponential theory in the given
approach. This shows that in genuinely nonlocal theories a new
type of one-loop divergences appear (divergences of large $N$).
Let us note that one can provide the finiteness of
the theory for each $N$ in a way we described above, and then
the limit $\,N\to\infty\,$ is well-defined.  For the  contributions
to the inverse of Newton constant this may be achieved by
adjusting the $\,\om_{N-1,\mathrm{GB}}\,$ and/or
$\,\om_{N,\mathrm{GB}}\,$ coefficients to provide zero result
for all $N$ in Eq. (\ref{divsGB}). These coefficients would
explicitly depend on $N$. For sensible nonlocal theory defined
in the limit $N\to\infty$ these coefficients should survive, but
the explicit evaluation of them shows that they are decaying like
$1/N$. In conclusion such generalized Gauss-Bonnet terms are
not present in the final action in the limit $N\to\infty$ of the
nonlocal theory. One may try to add standard killer operators
\cite{modestoLeslaw}. However, without explicit calculations
it is unclear whether this will be sufficient to provide an
UV-finite nonlocal theory.

 The expressions of the type $\om_{N-1,i}/\om_{N,i}$ appear in
 \eqref{EHdiv-simp}, where there is no mixing term. In full
 generality the diagonality in $i,j$ indices ($i,j=R$ or $C$) lets
 us to interpret such ratio in the limit $N\to\infty$ as a  d'Alembert
 definition of the convergence radius\footnote{Precisely speaking
 this is a convergence radius of a Taylor series expansion around
 the origin $z=0$.} of the complex function defined by the series
 $f_i(z)=\sum_{n=0}^\infty \om_{n,i} z^n$, that is
 \beq
 \rho_{f_i}=\lim_{N\to\infty} \frac{\om_{N-1,i}}{\om_{N,i}}\,.
 \eeq

 This result can be exported also for the similar interpretation of
 the divergence proportional to the cosmological constant. From
 this it is obvious that if any form factor like $\Phi$ or $\Psi$ is
 an entire function on the complex plane, then the new type of
 divergence will inevitably appear. However, this formula gives
 also the possibility of defining the standard divergences for
 theories in which $\Phi$ or $\Psi$ are other non-analytic real
 functions.

In the first part of this section we studied the simpler example
of  nonlocality that turns out to be sufficient to make convergent
scalar field theories in Euclidean signature. In Minkowskian
signature it is sufficient to replace
$\exp (- \Box/M^2)$ by $\exp P(\Box/M^2)$, where
$P(\Box/M^2)$ is a polynomial of even degree in its argument.
However, in a gravitational or a non-abelian gauge theory case
such nonlocality seems not suitable exactly because of the
gauge symmetry that forces the kinetic and interaction operators
to have the same ultraviolet scaling, which implies nonlocal
divergences. However, we can overcome this issue with a special
class of  nonlocal form factors that enjoy the property to be
asymptotically polynomial  in a region around the real axis \cite{kuzmin,Tomboulis-97,modesto,modestoLeslaw}.

Let us give here an explicit example of such function which is
supposed  to replace the exponential $\exp (- \Box/M^2)$
in (\ref{PhiPsi}),
\beq
e^{H(z)} = e^{\frac{1}{2} \left[ \Gamma
\left(0, p(z)^2 \right)
+\gamma_E  + \ln \left( p(z)^2 \right) \right] }
\quad
\mbox{for}\
\quad
z=\Box,
\quad
\label{exampleT}
\eeq
where $p(z)$ is a polynomial of degree $N$, $\gamma_E$ is the
Euler-Mascheroni constant and finally $\Gamma(a,x)$ is the
incomplete Gamma function. The crucial property of the function
in \eqref{exampleT} is that the divergent contributions to the beta
functions only depend on the local asymptotic polynomial $p(z)$,
while the full nonlocality manifests itself only in the finite
contributions to the quantum effective action. It is easy to see that
any correction to the UV polynomial is exponentially suppressed
\cite{modestoLeslaw}. Therefore, all results which were reported
for the polynomial case can be applied to this class of theories.

Let us note that the beta functions in the case under consideration
are one-loop exact and that two out of the total four relevant
divergent  contributions to the quantum effective action are
known. The last observation is that in the paper \cite{kuzmin} it
is stated that the beta functions for $R^2$ and $R_{\mu\nu}^2$
vanish, but we are unable to confirm or verify this statement.

%%%%%%%%% red

All the beta functions for local or nonlocal theories have been
computed using the heat-kernel expansion and Barvinsky-Vilkovisky
trace technology, therefore, in Euclidean space. The analytic
continuation to the Minkowski space can be implemented on the
basis of recent developments. We will deal with this subject further
in a future work.

%%%%%%%%%%%%%%%%%%%%%%%%%%%%%
%%%%%%%%%%%%%%%%%%%%%%%%%%%%%
%%%%%%%%%%%%%%%%%%%%%%%%%%%%%
\section{Renormalization group equations}
\label{V5}

Let us come back to the polynomial theory (\ref{gaction}) and
consider the case without killer terms, when the divergences are
given by the expressions (\ref{EHdiv-simp}) and (\ref{CCdivs}).
In this case one can construct the renormalization group
equations for the parameters $\om_{\rm EH}$ and
$\om_{\rm cc}$ of the classical action. The derivation of these
equations is almost a trivial task (see, e.g., \cite{book} for the
introduction), but for the sake of completeness we provide
the explanations here. The system of renormalization group
equations for the running coupling constants is defined by
relations
\beq
\frac{ d \alpha_i }{d t}
=
\beta_i \,,
\quad
{\rm with}
\quad
t =\ln \frac{\mu}{\mu_0} \,.
\eeq
Here $\alpha_i(t)$ are obtained by making the coupling constants
$\alpha_i$ scale-dependent  in the classical action. The
renormalized Lagrangian is
\beq
\mathcal{L}_{\rm ren}
&=&
\mathcal{L}(\alpha_i(t)) + \mathcal{L}_{\rm ct}
= \mathcal{L}( \alpha_i(t) )
\\
&-&
(Z_{\Lambda_{\rm cc}} -1) \frac{2 \Lambda_{\rm cc}}{16 \pi G}
- (Z_{G} -1) \frac{1}{16 \pi G} \, R
 + \dots \,,
\nonumber
\eeq
where $\mathcal{L}_{\rm ct} $ is a counterterm and $Z_{\alpha_i}$
denote renormalization constant of the couplings $\alpha_i$. Let us
refer to the further details in Appendix B.

The final results for the beta functions for $\,N \geqslant2$ are as
follows:
\beq
\beta_{G}
&=& \mu \frac{d}{d\mu}\,\Big( - \frac{1}{16 \pi G}\Big)
\\
&=&
-\,  \frac{1}{6(4\pi)^2}\,
\Big(\frac{5\om_{N-1,C}}{\om_{N,C}}
+ \frac{\om_{N-1,R}}{\om_{N,R}}\Big) ,
\label{runGom}
\nonumber
\\
\beta_{\rm cc}
&=&
\mu \frac{d}{d\mu}\Big( - \frac{\La_{\rm cc}}{8\pi G}\Big)
\\
&=&
\frac{1}{(4\pi)^2}
\Big(\frac{5 \omega _{N-2,C}}{\omega _{N,C}}
+\frac{\omega _{N-2,R}}{\omega _{N,R}}
\nonumber
\\
&-&
\frac{5 \omega _{N-1,C}^2}{2 \omega _{N,C}^2}
-\frac{\omega _{N-1,R}^2}{2 \omega_{N,R}^2} \Big) .
\label{runCCom}
\nonumber
\eeq
A pertinent observation is that for $N \geqslant 3$ these two
beta functions are exact, since higher loops would not provide
further contributions. The next question is that whether these
beta functions have physical sense. In other words, we have
to describe the situation when they may describe the running
of the gravitational and cosmological constants.

The physical sense of the running of the inverse Newton and 
cosmological constant is a complicated problem. It is known 
\cite{Barvinsky:2003kg} (see also \cite{apco}) that such a 
running does not correspond to the non-local $\Box$-dependent 
form factors in the loop corrections to the classical action. The 
reason is that in both cases such a form factors would act on 
a constant or produce a series of irrelevant (without boundaries) 
surface terms. These arguments mean that the running of $G$ and
$\La$ cannot be understood in a usual Quantum Field Theory 
sense, which is based on the scattering amplitudes on the flat
background. At the same time this does not forbid the running, 
because physically interesting gravitational  observables do 
not reduce to such amplitudes. The reader can consult
Refs.~\cite{CCrunning,PoImpo} for further references on this 
interesting issue, in the part of the cosmological constant running. 
Also, a kind of a regular procedure of the physically relevant 
scale identification \cite{Grunning} for the running of Newton 
constant was discussed in \cite{DomSte} and some other works. 

It is instructive to compare (\ref{runGom}) with the general
arguments based on covariance and dimensional considerations,
Eq.~(\ref{runningG}). For this end we have to identify what are
the masses in the theory under consideration. In general, the theory
of higher derivative quantum gravity (\ref{gaction}) possesses
many massive degrees of freedom, some of them ghost-like
and some normal \cite{highderi}. Then our Eq. (\ref{runGom})
is exactly (\ref{runningG}), where the dimensionless ratios
between different $\om$'s play the role of coupling constants.

The most interesting is the case when all massive states have
complex masses $m_i$. The theory may be consistent only if
these masses enter as complex conjugate pairs. This condition is
provided by a real classical action of the theory. But the same
condition applies also to the renormalization group improved
action. Therefore, in this case Eq. (\ref{runningG}) boils down
to the reduced version
\beq
\mu \frac{d}{d\mu}\,\frac{1}{G}
&=&
\sum\limits_{i} \al_i \,m^*_i  m_i
=
\sum\limits_{i} \al_i \,\left| m_i\right|^2\,,
\label{runGcomplex}
\eeq
where $\al_i$ are real coefficients depending on the dimensionless
ratios of the coefficients $\om$.

% \end{widetext}

As it was extensively discussed in \cite{CCrunning} and in the
review paper \cite{PoImpo}, it is not clear that both or any of
the renormalization group equations can be applied to cosmology
and astrophysics. The reason is that the diagrams which lead to
Eqs. (\ref{runGom}) and (\ref{runCCom}) include internal lines
of a massless graviton, but also of the number of massive states.
In the most interesting cases of unitary models of higher
derivative quantum gravity these massive poles are all complex
\cite{LM-Sh,LWQG}. Then we face an unsolved problem of what
remains from the effects of quantum gravity with massive modes
at low energy \cite{Gauss}. It is well-known that the contributions
of the loops of massive fields do decouple in the IR, in accordance
with the gravitational version of the Appelquist and Carazzone
theorem \cite{apco}.  However, this is a well-established result
only for the contributions of massive matter fields, when all
internal lines of the diagrams have the propagators with equal
masses. In the case of quantum gravity one has to deal with a
more complicated case of mixed loops, where some of  the
internal lines are massive and some are massless, or have much
lighter mass. Needless
to say that the situation becomes much more tricky in the case of
numerous complex masses. The analysis of this issue is certainly
very interesting, but it is beyond the scope of the present work. In
any case the expressions (\ref{runGom}) and (\ref{runCCom})
represent a good starting point, being a universal UV limits for
the physical beta functions under discussion.

%%%%%%%%%%%%%%%%%%%%%%%%%%%%%
%%%%%%%%%%%%%%%%%%%%%%%%%%%%%
%%%%%%%%%%%%%%%%%%%%%%%%%%%%%
\section{Conclusions}
\label{VI}

The one-loop calculations always represent one of the most
important elements in understanding new theories of quantum
gravity. These calculations are especially difficult in the case
of higher derivative models. In the present work, the one-loop
calculations in the super-renormalizable quantum gravity theory
has been extended to the term linear in scalar curvature. Together
with the previous derivation \cite{highderi} in the cosmological
constant sector, this enables us to obtain the closed system of
renormalization group equations  (\ref{runGom}) and
(\ref{runCCom}), which may be a basis for establishing the
physically relevant running of Newton and cosmological
constants in both high- and low-energy regimes.

In our opinion, the following two aspects of super-renormalizable
quantum gravity theories make the result especially relevant. First
of all, in the models with $\,N \geqslant 2\,$ the one-loop
beta functions are exact. Therefore, in (\ref{runGom}) and
(\ref{runCCom}) we have the first example of exact
beta functions in four-dimensional quantum gravity. Second,
some versions of these theories have only complex conjugate
pairs of massive ghost-like states, which turn out to be unitary
at tree-level \cite{LM-Sh,LWQG} and at perturbative level
\cite{CLOP}.
The results obtained here cover these cases and, therefore, can be
seen as a first example of beta functions in a class of consistent
models of perturbative quantum gravity.

One of the most remarkable features of the result (\ref{runGom}) is
its universality. For the models with $N \geqslant 3$ this is the first
example of an exact beta function for the Newton constant ever
calculated in quantum gravity models. Needless to say that
there are many approaches thet derive the non-perturbative
renormalization group flows in different models of quantum gravity.
The expressions (\ref{runGom}) and (\ref{runCCom}) can be used
as a test for these approaches - something that cannot be done with
(\ref{runCCom}) alone.

Last but not least, an important  aspect concerning the beta functions
(\ref{runGom}) and (\ref{runCCom}) is that they are based
on the Minimal Subtraction scheme of renormalization, and,
therefore, correspond to the running of $G^{-1}$ and $\rho_\La$
in the UV. The most interesting running of these parameters is the one in
the IR, and this subject opens a vast area for the future work,
where the universal expressions  $G^{-1}$ and $\rho_\La$ may
be used as tester for different approaches to the physical running
and consequent applications in cosmology and astrophysics.

%%%%%%%%%%%%%%%%%%%%%%%%%%%%%
%%%%%%%%%%%%%%%%%%%%%%%%%%%%%
%%%%%%%%%%%%%%%%%%%%%%%%%%%%%
\section*{Acknowledgements}

This work has been partially started during the visit of I.Sh. to
the group of Theoretical Physics in Fudan University. The kind
hospitality of this group is gratefully acknowledged.
I.Sh. is thankful to CNPq, FAPEMIG and ICTP for partial
support of his work. Part of the preparation of the manuscript
by I.Sh. was done during the workshop in Mainz, and I.Sh. is
grateful to the Mainz Institute for Theoretical Physics (MITP)
for the hospitality and partial support of his work during this
event.

%%%%%%%%%%%%%%%%%%%%%%%%%%%%%
%%%%%%%%%%%%%%%%%%%%%%%%%%%%%
%%%%%%%%%%%%%%%%%%%%%%%%%%%%%
\section*{Appendix A}

We here collect the list of bilinear expansions of
some relevant operators. Our attention will be restricted to the case
$N\geqslant 2$, when there is no need to expand Einstein-Hilbert
and cosmological terms. The expansions of the latter operators can
be easily found elsewhere \cite{avbar86,Gauss}. In the expressions
below the symmetrization in the pairs of indices $(\mu,\nu)$ and
$(\rho,\si)$ is supposed. We present the expansions valid on
maximally symmetric spacetimes, with $\La$ a constant parameter
characterizing the background curvature according to the formula
$\La=R/4$.

%%%%%%%%  WIDE
%%%%%%%%%%%%%%%%%%%%%%%%%%%%%%%%%%%
\begin{widetext}

We give the expression for the relevant part of the
bilinear form for the action \eqref{gaction}
\beq
 \frac{1}{\sqrt{|g|}}\,\frac{\de^2 S_{N}}{\de h_{\mu\nu}\,\de h_{\rho\si}}
&=&
\sum\limits_{n=0}^{N} S^{\prime\prime}_{(n)}{}^{\mu\nu,\,\rho\si},
\label{sum}
\eeq
where for $n\geqslant1$
\beq
%\hspace{-1.5cm}
S^{\prime\prime}_{(n)}{}^{\mu\nu,\,\rho\si}
&=&
g^{\mu\rho}g^{\nu\si}
\left[
\frac{2n-3}{3}\La \om_{n,C}+\frac{2n-3}{3}\La\om_{n,\text{GB}}
+\frac{1}{2}\om_{n,C}\square \right]\square^{n+1}
\nonumber
\\
&+&
g^{\mu\nu}g^{\rho\si}
\left[ -\frac{4n-5}{18} \La\om_{n,C}
- \frac{2n-1}{3}\La \om_{n,\text{GB}}-\frac{2(n-2)}{3}\La\om_{n,R}\right.
%\nonumber
%\\
%&&
-
\left. \left(
\frac16\,\om_{n,C}-\om_{n,R}\right)\square
\right]\square^{n+1}
\nonumber
\\
&+&
g^{\mu\nu}\na^{\rho}\na^{\si}
\left[
\frac{2n-1}{9}\, \La\om_{n,C}
+ \frac{2n-1}{3}\La\om_{n,\text{GB}}
+ \frac{8n+5}{3}\La\om_{n,R}\right.
%\nonumber
%\\
%&&
+ \left.
\left(
\frac16\,\om_{n,C}-\om_{n,R}\right)\square\right]\square^n
\nonumber
\\
&+&
g^{\rho\si}\na^{\mu}\na^{\nu}
\left[
\frac{2n-3}{9}\La\om_{n,C}
+ \frac{2n-1}{3}\La\om_{n,\text{GB}}
+ \frac{2n-3}{3}\La\om_{n,R}\right.
%\nonumber
%\\
%&&
+ \left.
\left(
\frac16\,\om_{n,C}-\om_{n,R}\right)\square\right]\square^n
\nonumber \\
%\eeq
%\beq
&+&
g^{\mu\rho}\na^{\nu}\na^{\si}
\left[
- \frac{4(n-1)}{3}\La\om_{n,C}
- \frac{2(2n-1)}{3}\La\om_{n,\text{GB}}
-\om_{n,C}\square \right]\square^{n}
\nonumber
\\
&+&
\na^{\mu}\na^{\nu}\na^{\rho}\na^{\si}
\left[
\frac{4}{9}n\La\om_{n,C}
-\frac{8}{3}n\La\om_{n,R}
+\frac{1}{3}\om_{n,C}\square
+\om_{n,R}\square \right]\square^{n-1}
+ {\cal O} \big(\La^2\big).\hspace{1cm}
\label{exp1}
\eeq

For the gauge-fixing term described in \eqref{gf}, \eqref{chi},
\eqref{weight} the expansion looks like
\beq
\frac{1}{\sqrt{|g|}}\,
\frac{\de^{2}S_{{\rm gf}}}{\de h_{\mu\nu}\de h_{\rho\si}}
&=&
\frac{1}{\sqrt{|g|}}\,
\frac{\de^2}{\de h_{\mu\nu}\de h_{\rho\si}}
\int\!d^{4}x\sqrt{|g|}\chi_{\mu}C^{\mu\nu}\chi_{\nu}
= \frac{2\beta\gamma}{\alpha}g^{\mu\nu}g^{\rho\sigma}
\left(\beta\square-\frac{2}{3}(N+1)\Lambda+\beta N\Lambda\right)\square^{N+1}
\nonumber
\\
&+&
g^{\rho\sigma}\nabla^{\mu}\nabla^{\nu}
\left(-\frac{2\gamma\beta}{\alpha}(\square+N\Lambda)
+\frac{4}{3\alpha}(1+\gamma N)\Lambda\right)\square^N
+ \frac{2\gamma\beta}{\alpha}g^{\mu\nu}\nabla^{\rho}\nabla^{\sigma}
\left(-\square+\frac{5N+8}{3}\Lambda\right)\square^{N}
\nonumber
\\
&+&
\frac{2}{\alpha}g^{\mu\rho}\nabla^{\nu}\nabla^{\sigma}
\left(\square-\frac{5N+8}{3}\Lambda\right)\square^{N}
+ \frac{2(\gamma-1)}{\alpha}
\nabla^{\mu}\nabla^{\nu}\nabla^{\rho}\nabla^{\sigma}
\left(\square-\frac{5}{3}N\Lambda\right)\square^{N-1}
+ O\left(\La^{2}\right) .
\label{expGF}
\eeq
This expression corresponds to arbitrary values of gauge-fixing
parameters, but for the practical calculations we used (\ref{mini}).
Let us stress that the final result is independent on the choice of
$\,\al$, $\,\be\,$ and $\,\ga$. The expressions \eqref{exp1} and
\eqref{expGF} were obtained from the self-adjoint forms of the
variational derivative operators after the commutations of
derivatives and expansions in powers of $\La$.

Now we want to perform an additional check. Since from the beginning
of our computation, we used "the best" basis in \eqref{gaction},
where the contributions to the propagators from particular terms diagonalize,
we were not able to do the check of the generalized GB term directly from
the combination of terms with Ricci scalar, tensor and Riemann tensor.
Instead we had separately terms in the action with Ricci scalars, Weyl tensors,
and this generalized GB term. Now we may investigate the effect of the
generalized GB term on flat spacetime. Its second variation should vanish
(after integration by parts, or equivalently in the self-adjoint form of the
second variational derivative). One easily checks that indeed for our variations
the contribution from the generalized GB is clearly zero on the flat spacetime.
In the formula \eqref{exp1} there are no terms proportional to
$\omega_{n,{\rm GB}}$, which would survive on flat spacetime, where
$\Lambda\to0$. The only terms with $\omega_{n,{\rm GB}}$ are
proportional to $\Lambda$. So we proved that in any dimension the
contribution to graviton's propagator around flat spacetime from a
generalized Gauss-Bonnet term for any $n$ is zero.

In the rest of this Appendix we present the elements of the  tensor
$H'$ from Eq.~(\ref{Hprime}). In all formulas below the
symmetrization in the pair of indices $(\ka,\la)$ is
assumed. For the calculation of
the linear in curvature part of the effective action it is sufficient
to consider $V$ and $U$ tensors. According to the notation
introduced in formula \eqref{expashort} we can write
\beq
V_{\ka\la}{}^{\rho\si,\,\al\be\ga\de}
&=&
{\tilde V}_{(0)\,\ka\la}{}^{\rho\si,\,\al\be\ga\de}
+ \La {\tilde V}_{(1)\,\ka\la}{}^{\rho\si,\,\al\be\ga\de},
\\
\nonumber
\\
U_{\ka\la}{}^{\rho\si,\,\al\be\ga\de}
&=&
{\tilde U}_{(0)\,\ka\la}{}^{\rho\si,\,\al\be\ga\de}+\La\, {\tilde U}_{(1)\,\ka\la}{}^{\rho\si,\,\al\be\ga\de}+\La^2\, {\tilde U}_{(2)\,\ka\la}{}^{\rho\si,\,\al\be\ga\de}\,.
\eeq

Due to dimensional arguments in the ${\tilde U}_{(0)}$ tensor we
would have the appearance of coefficients of the type
$\om_{N-2,i}/\om_{N,j}$ (where $i,j=R,C,$ or ${\rm GB}$),
while we know from considerations in the main text that this tensor
would not contribute to divergences proportional to the linear term
in scalar curvature. Similarly, we know that in the ${\tilde U}_{(2)}$
tensor we find coefficients $\om_{N,i}/\om_{N,j}$, and again they
cannot contribute to the desired divergence because we know from
\eqref{expa} that for divergences we can only take the trace of the
$U$ tensor (with no interference terms). Hence we neglect writing
the form of these parts of the tensor $U$ and we concentrate below
only on the part ${\tilde U}_{(1)}$ that gives contribution.

The list of the corresponding expressions is as follows
\beq
{\tilde V}_{(0)\,\ka\la}{}^{\rho\si,\,\al\be\ga\de}
&=&
\Big[\de_{\ka}^{\rho}\de_{\la}^{\si}g^{\al\be}g^{\ga\de}
- \frac13\,g_{\ka\la}g^{\rho\si}g^{\al\be}g^{\ga\de}
+ \frac13\,g^{\rho\si}\de^\al_{\ka}\de^\be_{\la}g^{\ga\de}
\nonumber
\\
&+&
\frac13\,g_{\ka\la}g^{\al\rho}g^{\be\si}g^{\ga\de}
%% \frac{\om_{N-1,C}}{\om_{N,C}}
+ %% \left[
\frac23\de^\al_{\ka}\de^\be_{\la}g^{\ga\rho}g^{\de\si}
%% \frac{\om_{N-1,C}}{\om_{N,C}}
- 2\de_{\ka}^{\rho}\de^\al_{\la}g^{\be\si}g^{\ga\de}\Big]
\frac{\om_{N-1,C}}{\om_{N,C}}
\\
&+&
2\left[\de^\al_{\ka}\de^\be_{\la}g^{\ga\rho}g^{\de\si}
%% \frac{\om_{N-1,R}}{\om_{N,C}}
- g^{\rho\si}\de^\al_{\ka}\de^\be_{\la}g^{\ga\de}\right]
\frac{\om_{N-1,R}}{\om_{N,C}}
+ \left[g_{\ka\la}g^{\rho\si}g^{\al\be}g^{\ga\de}
%% \frac{\om_{N-1,R}}{3\om_{N,R}}
- g_{\ka\la}g^{\al\rho}g^{\be\si}g^{\ga\de}\right]
\frac{\om_{N-1,R}}{3\om_{N,R}}\,,
\nonumber
\label{V0}
\eeq
%%%%%%%%%%%%%%%%%%%%%%%%%%%
%%%%%%%%%%%%%%%%%%%%%%%%%%%
\beq
{\tilde V}_{(1)\,\ka\la}{}^{\rho\si,\,\al\be\ga\de}{}
&=&
\frac{2(2N-3)}{3}\de_{\ka}^{\rho}\de_{\la}^{\si}g^{\al\be}g^{\ga\de}
-\frac{2(3N-5)}{9}g_{\ka\la}g^{\rho\si}g^{\al\be}g^{\ga\de}
+ \frac{3N+2}{3}g^{\rho\si}\de^\al_{\ka}\de^\be_{\la}g^{\ga\de}
\nonumber
\\
&+&
2N\de^\al_{\ka}\de^\be_{\la}g^{\ga\rho}g^{\de\si}
%% \nonumber \\ &+&
+ 2\Big[ (N-1)g^{\rho\si}\de^\al_{\ka}\de^\be_{\la} g^{\ga\de}
%% \frac{\om_{N,R}}{\om_{N,C}}
- N\de^\al_{\ka}\de^\be_{\la}
g^{\ga\rho}g^{\de\si}
+ \frac13 g_{\ka\la}
A^{\al\rho,\,\be\si}
g^{\ga\de} \Big]\frac{\om_{N,R}}{\om_{N,C}}
\nonumber
\\
&+&
\frac{2(2N-1)}{3}
\Big[
\de_{\ka}^{\rho}\de_{\la}^{\si}g^{\al\be}g^{\ga\de}
+ \frac13\,g_{\ka\la} A^{\al\rho,\,\be\si} g^{\ga\de}+
\de^\al_{\ka}
\left(g^{\rho\si}\de^\be_{\la}-2\de_{\la}^{\rho}g^{\be\si}\right)
g^{\ga\de}\Big]\, \frac{\om_{N,\text{GB}}}{\om_{N,C}}
\nonumber
\\
&+&
\frac{2(2N-1)}{27}g_{\ka\la}
A^{\al\rho,\,\be\si}
g^{\ga\de}
\frac{\om_{N,\text{GB}}}{\om_{N,R}}
- \frac{2(9N+4)}{3}\de_{\ka}^{\rho}\de^\al_{\la}g^{\be\si}g^{\ga\de}
+ \frac{2(6N+1)}{9}g_{\ka\la}g^{\al\rho}g^{\be\si}g^{\ga\de},
\label{V1}
\eeq
where
\beq
A^{\al\rho,\,\be\si} =
g^{\al\rho}g^{\be\si}-g^{\rho\si}g^{\al\be}\,.
\nonumber
\eeq
%%%%%%%%%%%%%%%%%%%%%%%%%%%%%%
%%%%%%%%%%%%%%%%%%%%%%%%%%%%%%
%%%%%%%%%%%%%%%%%%%%%%%%%%%%%%
Finally,
\beq
{\tilde U}_{(1)\,\ka\la}{}^{\rho\si,\,\al\be\ga\de}
&=&
2\de_{\ka}^{\rho}\de_{\la}^{\si}g^{\al\be}g^{\ga\de}
\left(\frac{2N-5}{3} \frac{\om_{N-1,C}}{\om_{N,C}}
+\frac{2N-3}{3} \frac{\om_{N-1,\text{GB}}}{\om_{N,C}}\right)
\nonumber
\\
\nonumber
\\
&+&
g_{\ka\la}g^{\rho\si}g^{\al\be}g^{\ga\de}\left(
-\frac{12N-31}{27} \frac{\om_{N-1,C}}{\om_{N,C}}
-\frac{2(2N-3)}{9} \frac{\om_{N-1,\text{GB}}}{\om_{N,C}}
-\frac{2}{9} \frac{\om_{N-1,R}}{\om_{N,C}}
\right)
\nonumber
\\
\nonumber
\\
&+&
g_{\ka\la}g^{\rho\si}g^{\al\be}g^{\ga\de}
\left(-\frac{2}{81} \frac{\om_{N-1,C}}{\om_{N,R}}
-\frac{2(2N-3)}{27} \frac{\om_{N-1,\text{GB}}}{\om_{N,R}}
- \frac{6N-19}{27} \frac{\om_{N-1,R}}{\om_{N,R}}\right)
\nonumber
\\
\nonumber
\\
&+&
4\de_{\ka}^{\rho}\de^\al_{\la}g^{\be\si}g^{\ga\de}
\left(
-\frac{2(N-2)}{3} \frac{\om_{N-1,C}}{\om_{N,C}}
-\frac{2N-3}{3} \frac{\om_{N-1,\text{GB}}}{\om_{N,C}}\right)
\nonumber
\\
\nonumber
\\
&+&
2g_{\ka\la}g^{\al\rho}g^{\be\si}g^{\ga\de}
\left(
\frac{6N-17}{27} \frac{\om_{N-1,C}}{\om_{N,C}}
+\frac{2N-3}{9} \frac{\om_{N-1,\text{GB}}}{\om_{N,C}}
-\frac{5}{9} \frac{\om_{N-1,R}}{\om_{N,C}}
\right)
\nonumber
\\
\nonumber
\\
&+&
2g_{\ka\la}g^{\al\rho}g^{\be\si}g^{\ga\de}
\left(
\frac{4}{81} \frac{\om_{N-1,C}}{\om_{N,R}}
+\frac{2N-3}{27} \frac{\om_{N-1,\text{GB}}}{\om_{N,R}}
+\frac{2(6N-1)}{27} \frac{\om_{N-1,R}}{\om_{N,R}}
\right)
\nonumber
\\
\nonumber
\\
&+&
2 g^{\rho\si}\de^\al_{\ka}\de^\be_{\la}g^{\ga\de}
\left(
\frac{2N-5}{9} \frac{\om_{N-1,C}}{\om_{N,C}}
+\frac{2N-3}{3}\frac{\om_{N-1,\text{GB}}}{\om_{N,C}}
+\frac{2N-5}{3} \frac{\om_{N-1,R}}{\om_{N,C}}
\right)
\nonumber
\label{Ublock}
\\
\nonumber
\\
&+&
8\de^\al_{\ka}\de^\be_{\la}g^{\ga\rho}g^{\de\si}
\left(
\frac{N-1}{9} \frac{\om_{N-1,C}}{\om_{N,C}}
-\frac{2(N-1)}{3} \frac{\om_{N-1,R}}{\om_{N,C}}\right),
\eeq
completing the list of necessary formulas.

%%%%%%%%%%%%%%%%%%%%%%%%%%%
%%%%%%%%%%%%%%%%%%%%%%%%%%%
\section*{Appendix B}

We here remind the standard definitions from the review paper
\cite{avbar86}.

The divergent contribution to the quantum effective action is given by:
\beq
\Gamma_{\rm div}
&=& \frac{i}{2} {\rm Det}{H'}
=  \frac{i}{2} {\rm Tr} \ln H'
=  \frac{i}{2} ( i \ln L^2) \!\int\! d^4 x \sqrt{|g|} \left( \beta_C C^2
+ \beta_R R^2 + \beta_{G} R +  \beta_{\Lambda_{\rm cc} }\right)
\nonumber
\\
&=&
% \hspace{0.75cm}
 -\, \frac{1}{2}   \ln L^2 \!\int\! d^4 x \sqrt{|g|} \left( \beta_C C^2
 + \beta_R R^2 + \beta_{G} R +  \beta_{\Lambda_{\rm cc} }\right),
 \label{GammaDiv}
\eeq
where
\beq
\ln L^2 = \frac{1}{2 - \omega},
\qquad
\omega = 2 - 0^+ = 2 - \vp   = \frac{n}{2}
\quad \Longrightarrow \quad
\ln L^2 = \frac{1}{\vp}>0 ,
\eeq
where $L = \Lambda/\mu$, \ $\Lambda$ is the cut-off scale and $\mu$
the renormalization scale.

Moreover,
\beq
n = 2 \omega = 2(2 - 0^+) = 2 ( 2 - \vp) = 4 - 2 \vp
\quad \Longrightarrow \quad
\ln L^2 = \frac{2}{ 4 - n} = \frac{1}{\vp} \,,
\eeq
where $n$ is the dimensionality of spacetime.
Then eq. (\ref{GammaDiv}) turns into
\beq
&& \Gamma_{\rm div} =  -  \frac{ 1}{2 \vp}
\!\int\! d^4 x \sqrt{|g|} \left( \beta_C C^2
+ \beta_R R^2 + \beta_{G} R +  \beta_{\Lambda_{\rm cc} }\right) \, .
\eeq
Finally, we give here the divergent contributions to the trace of the
logarithm of $H'$ operator for the following different values of $N$
in dimensional regularization scheme.

For $N\geqslant2$:
\beq
{\rm Tr}\ln H'|_{\rm div}
&=&
\frac{i\ln L^{2}}{16\pi^{2}}\!\int\! d^{4}x\sqrt{|g|}(-1)
\bigg\{
\frac{5 \om_{N-1,C}}{6\om_{N,C}}
+ \frac{\om_{N-1,R}}{6\om_{N,R}}
+ (2N-1)\left[
\frac{5 \,\om_{N-1,C}}{6\om^2_{N,C}}
- \frac{\om_{N-1,R}\,}{18\om^2_{N,R}}
\right]
\,\om_{N,\mathrm{GB}}
\nonumber
\\
&+&
(2N-3)\left[
\frac{5}{6\om_{N,C}}
- \frac{1}{18\om_{N,R}}
\right]\,\om_{N-1,\mathrm{GB}}
\bigg\}R\,\,+\,O\left({\cal R}^0,{\cal R}^2\right)\, .
\eeq

For $N=1$:
\beq
{\rm Tr}\ln H'|_{\rm div}
&=&
\frac{-i\ln L^{2}}{16\pi^{2}}\!\int\! d^{4}x\sqrt{|g|}
\left\{
\frac{5\om_{0,C}}{6\om_{1,C}}
+ \frac{\om_{0,R}}{2\om_{1,R}}
- \frac{5\om_{0,R}}{\om_{1,C}}
+ \left(
\frac{5\om_{0,C}}{6\om_{1,C}^2}
- \frac{\om_{0,R}}{18\om_{1,R}^2}
\right) \om_{1,\mathrm{GB}}
\right\}R + O\left({\cal R}^0,{\cal R}^2\right).
%% \nonumber
\eeq

For $N=0$:
\beq
{\rm Tr}\ln H'|_{\rm div}
&=&
\frac{i\ln L^{2}}{16\pi^{2}}\!\int\! d^{4}x\sqrt{|g|}
\left(-\frac{5 \omega_{\rm EH}\, \omega_R}{2 \omega_C^2}
-\frac{13 \omega_{\rm EH}}{12
 \omega_C}+\frac{\omega_{\rm EH}}{12 \omega_R}\right)R
+O\left({\cal R}^0,{\cal R}^2\right) \,.
\eeq

And for $N\geqslant2$ (divergence of the cosmological constant term):
\beq
\Tr \ln H'|_{\rm div}
&=&
\frac{i\ln L^{2}}{16\pi^{2}}\!\int\! d^{4}x\sqrt{|g|}
\left(\frac{5 \omega _{N-2,C}}{\omega _{N,C}}
+\frac{\omega _{N-2,R}}{\omega _{N,R}}
-\frac{5 \omega _{N-1,C}^2}{2 \omega _{N,C}^2}
-\frac{\omega _{N-1,R}^2}{2 \omega_{N,R}^2}\right)
\,\,+\,O\left({\cal R}^1\right) \,,
\eeq
where by $\cal R$ above we mean a general gravitational curvature
tensor (Ricci scalar, tensor or Riemann tensor).

%%        WIDE
\end{widetext}

%%%%%%%%%%%%%%%%%%%%%%%%%%%%
%%%%%%%%%%%%%%%%%%%%%%%%%%%%%
%%%%%%%%%%%%%%%%%%%%%%%%%%%%%
\section*{Appendix C}

Let us comment on the interesting issue of stability in the
theory with higher derivatives and complex poles.
%% \footnote{We acknowledge the stimulating
%% requirement of one of the anonymous referees
%% to include the discussion on this issue.}.

The problem of stability of classical solutions is the most difficult
one for the theories with higher derivatives. There are different
approaches to this important problem. Let us describe the possibility
which takes place within the Lee-Wick construction \cite{LW}.
Indeed, as far as one accepts this approach, the problem is
greatly alleviated.
The Lee-Wick proposal \cite{LW} consists of removing
the instabilities by requiring, as a boundary condition, that the
growing modes do not appear on-shell (see also the discussion
below about tree-level and perturbative unitarity).
As a consequence of eliminating the exponentially growing states,
acausal effects occur, just as in the classical theory of electron, when
runaway modes have to be eliminated. This is not surprising: a future
boundary condition has been imposed and the initial state must be
such as to eliminate any growing modes that would otherwise be
created. This topic has been extensively discussed in the literature;
the results may look non-conventional but do not appear to conflict
with any experimental observations, nor do any paradoxes or
inconsistencies arise.

Furthermore, the linear stability analysis comes together with the
unitarity issue because both have to do with the $S$-matrix at
perturbative level. The former at tree-level while the latter at any
order in the loop
expansion. The problem is to define a unitary $S$-matrix, but
fortunately, this was done by Lee and Wick. Indeed, the evolution
operator $U(t, t_0)$ is not unitary and, moreover, it becomes
unbounded as $t \rightarrow \pm \infty$. However, the physical
states are defined so as to exclude the complex-mass particles from
the asymptotic states and contain only real-mass particles. The
$S$-matrix is obtained by subtracting from $U$ the terms that
grow exponentially, and is shown to be unitary \cite{LW}.

Equivalently, the $S$-matrix is obviously unitary in the whole
Hilbert space of real and complex states, then, in order to
prove the unitarity of the $S$-matrix in the physical subspace of
real states, we have only to prove that if the initial state is physical, then the final state is physical too. The key
point here is that the energy conservation comes together with the $S$-matrix,
and, therefore, if the initial state has real energy also the final state
has real energy, otherwise the energy would get an imaginary part. The orthogonality between real and complex energy states is satisfied on both sides of the unitarity equation
$T - T^\dagger = i \, T^\dagger T$,
which comes from $S^\dagger S = 1$ and $S= 1 + i \, T$.
It is easy to see that the $S$-matrix is unitary in the physical subspace of real states (in the minimal Lee-Wick gravitational theory we only have the graviton state). The complex-mass
particles appear only in virtual states which never go on-shell and never
become asymptotic states. We can have complex conjugate ghost pairs with a positive norm and real energy in a special Lorentz frame, but their energy is complex in general Lorentz frames. These states have
to be projected out as well but the argument above does not change
because such states are in the Hilbert space of complex ghosts. Indeed, they are tensor products of complex states.
Finally, we can get negative norm states combining complex ghosts' states but again these states are complex and, therefore, outside the physical subspace.

The classical stability is the most trivial application of the unitarity
analysis namely, we use the tree-level $S$-matrix scattering amplitude.
Of course, the action is not sufficient to define the theory and we
have to add the Lee-Wick prescription to get something meaningful.
Naively, there are runaway solutions, but when the Hilbert space of
the asymptotic states is properly projected there are no such instabilities.
A much more difficult task is to prove the perturbative unitarity.
However, the latter was proved by Cutkosky, Landshoff, Olive, and
Polkinghorne (CLOP) in Ref.~\cite{CLOP} as it is shortly summarized
in the paper \cite{LM-Sh} of our group. Finally, the CLOP prescription allows for the Wick
rotation (see also \cite{Wise}), which justifies our computation also in
the Minkowski space. Other forms of ``non-analytic'' continuation from
Minkowskian to Euclidean signature can be also found in the literature. % \cite{A1,A2,A3}.
These results make the theory published in \cite{LM-Sh} a very good proposal for quantum gravity \cite{LM-Sh,LWQG}.

At classical level, another well-defined prescription, similar to the Lee-Wick one, to
remove the runaway solutions  was proposed in
Ref.~\cite{Barnaby:2007ve}.

It is impossible to say that the solutions of the stability problem in
higher derivative models is ultimately achieved by the considerations
presented above. However, at the present-day level of understanding
these considerations look useful and may be instructive in
constructing a consistent theory of quantum gravity. Another
approach to the problem of stability has been recently
discussed in \cite{PP-17}.

%%%%%%%%%%%%%%%%%%%%%%%%%%%%%
%%%%%%%%%%%%%%%%%%%%%%%%%%%%%
%%%%%%%%%%%%%%%%%%%%%%%%%%%%%
\section*{Appendix D}

As we have explained in the main part of the paper, the beta
functions for the cases $N \geqslant 1$ are universal in the sense
they are independent on the choice of gauge-fixing and/or
parametrization of a quantum field. This feature is in a sharp
contrast with the situation in quantum gravity based on
Einstein's general relativity (see recent discussion in \cite{DG-QG}
and further references therein) and four-derivative quantum gravity.

One can naturally ask whether some other type of ambiguity can
be detected in the super-renormalizable model of quantum gravity.
Let us consider, for instance, the transformation of the metric
\beq
g_{\mu\nu} \,\longrightarrow\, g'_{\mu\nu}
\,=\,g_{\mu\nu} + {\cal X}_{\mu\nu}.
\label{X1}
\eeq
The question is whether such a transformation can modify the
beta functions\footnote{We acknowledge the stimulating requirement
of one of the anonymous referees to include the discussion on this issue.}.
The transformation of this sort has been discussed long time ago in
relation to renormalization of quantum general relativity \cite{KTT}
(see also more general discussion in \cite{VT-1981}), but it is
useful to analyze it in the essentially more general context.

It is evident that one can apply (\ref{X1}) in two different ways. The
first one is not to assume any relation between ${\cal X}_{\mu\nu}$ and the
quantum corrections. In this case, it is clear that choosing the trace
of the tensor ${\cal X}_{\mu\nu}$ in a special way one can modify
all terms in the action, except the cosmological constant term (in a
case we do not like to modify it). In
order to understand this, let us present the total Lagrangian of the
theory (both the gravitational and matter parts) in the form
\beq
L_{\rm tot} \,=\, \om_{\rm cc} + L_A.
\label{X2}
\eeq
The total classical action is
\beq
S_{\rm tot}\,=\,\int\! d^4 x \sqrt{|g|} \,L_{\rm tot}.
\label{X4}
\eeq

Under the replacement (\ref{X1}) the determinant in (\ref{X4})
transforms according to the Liouville's formula
\beq
g^\prime &=& g
\times \exp\Big\{{\cal X} - \frac12 {\cal X}_{\mu\nu}^2
+ \frac13 {\cal X}_{\mu\nu}^3
- \frac14 {\cal X}_{\mu\nu}^4 + \dots\Big\},
\label{Xe}
\eeq
where the contractions are done with the metric, e.g.,
\ ${\cal X}=g^{\mu\nu}{\cal X}_{\mu\nu}$, \
$\,{\cal X}_{\mu\nu}^2
= g^{\mu\al}g^{\nu\be}{\cal X}_{\mu\nu}{\cal X}_{\al\be}$,
etc. According to the last formulas, the choice
\beq
{\cal X}_{\mu\nu} \,=\,-\, \frac{L_A}{2\,\om_{\rm cc}}\,g_{\mu\nu}
\label{X3}
\eeq
eliminates all the terms in $S_{\rm tot}(g')$, which are linear in $L_A$,
these terms being gravitational or matter Lagrangians.

After the operation (\ref{X1}), in terms of the metric $g_{\mu\nu}$
the theory changes dramatically. At first order in $L_A$ all the
terms in the action disappear and there is only the cosmological
constant remaining. Then the next order terms ${O}(L_A^2)$
can be eliminated
by further tuning of the tensor ${\cal X}_{\mu\nu}$ which becomes
an infinite series in $L_A$. \ The proof that it is possible to do this
non-perturbatively is left to the interested reader as an exercise.

At this level, the operation (\ref{X1}) has nothing to do with quantum
corrections, it is a purely classical change of the metric variable which kills
all terms in the action except the $\om_{\rm cc}$, namely the cosmological
constant term. One can, of course, make changes in the operation
(\ref{X1}), e.g., kill only purely gravitational terms, or purely matter
terms, or make another selection. In this case there will be obvious
complications, but let us skip this part for the sake of brevity.

The application of the operation (\ref{X1}) at the quantum level
opens new possibilities for further speculations. In this case the
classical action (\ref{X4}) is replaced by the quantum analog called
effective action (see \cite{book} for the details in curved
spacetime and in quantum gravity),
\beq
\Ga\,=\,S_{\rm tot}\,+\,{\bar \Ga}, \qquad
{\bar \Ga} = \sum\limits_{k=1}^{\infty} \hbar^k{\bar \Ga}^{(k)},
\label{X5}
\eeq
where the last sum corresponds to the loop expansion. One can
link the operation (\ref{X1}) with the parameter of the loop
expansion $\hbar$. It is clear that taking \
${\cal X}_{\mu\nu} = { O}(\hbar)$ one can, for example,
kill completely or partially, or just modify the one-loop
corrections, while the classical terms $L_A$ will not be
affected at all. Furthermore, taking ${\cal X}_{\mu\nu}$ to be an
infinite series in $\hbar$, one can arbitrarily (or almost so) modify
higher loop corrections too. One can perform a transformation
\eqref{X1} with the loop parameter $\hbar$ and then all the
beta functions will be affected (can be canceled, for instance),
but this would be as artificial as in the general classical case
considered above.

One can make speculations on the basis of the operation (\ref{X1}).
For instance, in quantum gravity one can start with the action with
four derivatives, or with the super-renormalizable action
(\ref{gaction}), calculate quantum corrections and then postulate
that the {\it physical} observable metric is not the original one
$g_{\mu\nu}$, but some other one, obtained by the operation (\ref{X1}),
such that the resulting theory has no ghosts. One can see an example
of this procedure described in the recent paper Ref.~\cite{Slovick}.
This approach to quantum gravity
has only three problems: ({\it\!i}) the procedure of changing the field
in an arbitrary way is not a change of variables (classical or quantum)
and does not belong to the conventional quantum field theory;
\ ({\it\!ii}) in some cases it requires infinitely precise fine-tuning of
infinitely many parameters, as it was recently discussed in
\cite{ABSh1} in relation to string theory and Zwiebach transformation
\cite{zwie};
\ ({\it\!iii}) there are serious ambiguities
in the resulting action since there are many higher derivative terms
which do not produce ghosts, and which can be eliminated or not,
depending on our will \cite{maroto}.

As far as our interest in the present paper is related
to the one-loop beta functions for the Newton constant $G$ and the
cosmological constant $\La_{\rm cc}$, we can restrict our attention to
the corresponding terms only\footnote{The interested reader can
easily make generalizations to other terms in the gravitational or
matter fields actions and explore the ambiguity of the corresponding
beta functions.}. Since we will deal with the
${\overline {\rm MS}}$-scheme of renormalization, we can also
restrict our attention to the corresponding version of the
renormalization group in curved space, as described in a
pedagogical form in \cite{book}. In order to simplify things as
much as possible, consider the theory (\ref{gaction}) with $N \geqslant 3$,
such that one-loop beta functions are exact and non-zero only for
the parameters
$\,\om_i=\big(\om_{\rm EH},\,\om_{\rm cc},\,\om_{0,R},\,
\om_{0,C},\,\om_{0,\mathrm{GB}}\big)$.

Then the simplest procedure is to take the quantum effective action
as the renormalization group improved classical action
\beq
\Ga \,\approx\, S_{\rm tot}^{\rm RG}
\,=\,S_{\rm tot}\big(\om_i(t)\big),
\label{X7}
\eeq
where in the one-loop approximation the corresponding expressions are
\beq
&& \om_1(t)\,=\,\om_{\rm EH}(t)
\,=\, \om_{\rm EH}(0) \,+\, \beta_{G}\,t,
\nonumber
\\
&& \om_2(t)\,=\,\om_{\rm cc}(t)
\,=\, \om_{\rm cc}(0) \,+\, \beta_{\rm cc}\,t,
\nonumber
\\
&&
\om_{3,4,5}(t) \,=\, \om_{3,4,5}(0) \,+\,\be_{3,4,5}\,t.
\label{X8}
\eeq
Actually, the last three beta functions were not calculated yet, but
we do not need them now.  In these formulas the ``RG time'' $t$
can be understood in two different ways. Taking the renormalization
group in terms of the renormalization parameter $\mu$, we can define
$t=\ln\big( \mu/\mu_0\big)$, such that the value $t=0$ corresponds
to the reference scale $\mu_0$. On the other hand, it is always
necessary to relate an artificial parameter $\mu$ to some physical
quantity. In curved space the unique well-developed identification
is to associate $t$ with the parameter of the global rescaling of the
metric \cite{nelpan82} (see also \cite{buch-84,book} for a consistent
treatment and full details),
\beq
g_{\mu\nu}\,\longrightarrow\, g_{\mu\nu}e^{2t}\,.
\label{X9}
\eeq
Let us use this fact as a hint to explore how the operation (\ref{X1})
can affect the beta functions $\beta_{G}$ and $\beta_{\rm cc}$.

As far as the beta functions are related to quantum corrections, we can
regard both  $\beta_{G}$ and $\beta_{\rm cc}$ as ${O}(\hbar)$.
Then the first two relations in  (\ref{X8}) can be recast as
\beq
\om_{\rm EH}(t)
= \om_{\rm EH}(0) \exp
\Big\{ \frac{\beta_G\,t}{ \om_{\rm EH}(0)}\Big\},
\nonumber
\\
\om_{\rm cc}(t)
= \om_{\rm cc}(0)
\exp \Big\{ \frac{\beta_{\rm cc}\,t}{ \om_{\rm cc}(0)}\Big\}.
\label{X10}
\eeq
Then the renormalization-group improved classical action (\ref{X7})
becomes
\beq
S_{\rm tot}^{\rm RG}
\,=\,
\int \!d^4x\sqrt{|g|}\,\big\{ \om_{\rm cc}(t)
\,+\,\om_{\rm EH}(t)\,R+O({\cal R}^2)
%% +{ O}(\hbar^2) Leslaw, there is no such term in the given conditions
\big\}.
\label{X11}
\eeq
Consider a very particular version of  (\ref{X1}),
\beq
g_{\mu\nu}\,\longrightarrow\,g'_{\mu\nu}\,=\,e^{2\al t}g_{\mu\nu},
\label{X12}
\eeq
where $\al$ is an arbitrary parameter of the first order in $\hbar$.
Replacing (\ref{X12}) into
(\ref{X11}), after a small algebra we find that the beta functions
get modified according to
\beq
&&
\beta_{\rm cc}\,\longrightarrow\,\be'_{\rm cc}
\,=\,\beta_{\rm cc} \,+\, 4\al\, \om_{\rm cc}(0),
\nonumber
\\
&&
\be_{G}\,\longrightarrow\,\be'_{G}
\,=\,\beta_{G} \,+\, 2\al\, \om_{\rm EH}(0).
\label{X13}
\eeq
Since we are interested only in the first order in $\hbar$, it is
possible to trade $\om_{\rm cc}(0)$ and $\om_{\rm EH}(0)$
in expressions (\ref{X13}) to fully $t$-dependent $\om_{\rm cc}(t)$
and $\om_{\rm EH}(t)$. And then it is clear that the quantity which
has $\al$-independent derivative with respect to $t$ is the
dimensionless ratio
\beq
\frac{\om_{\rm cc}(t)}{\om^2_{\rm EH}(t)}\,\sim\,G\La_{\rm cc}.
\label{X14}
\eeq
By coincidence, this is the same combination that does not depend
on the choice of gauge-fixing in four-derivative gravity and appears in
the on-shell renormalization group in perturbative quantum Einstein's
general relativity \cite{frts82} (see also \cite{avbar86} and \cite{a} for
further details in the first case and \cite{DG-QG} in the last one).

Does it mean that the ambiguity related to (\ref{X12}) and the one
related to the choice of the gauge-fixing or parametrization of
quantum fields are somehow related? In
our opinion the answer is negative. As we have argued above, the
transformation (\ref{X1}) can produce much more significant
changes in the theory at both classical and quantum level than the
change of the gauge-fixing. For instance, it is an easy exercise to
show that (\ref{X1}) can change the sign of the beta function for
the running charges in QED or in QCD. The corresponding
beta functions do not depend on the gauge-fixing and do not have
any relation to gravity or metric, so this operation is obviously
senseless. In general, the operation (\ref{X1}) and its particular
version (\ref{X12}) are not related to quantum field theory. These
operations with the {\it background} quantities  are not related to
legitimate changes of quantum variables in the path integral
formalism. The reason is that the corresponding transformation
 should be done for the background metric and is not the change
 of quantum variables. One can say that the transformations
 \eqref{X1} should be ``forbidden'' if we intend to explore the
 quantum field theory formalism. All in all, (\ref{X1}) is a technical
 trick which is not related to quantization at all.

%%%%%%%%%%%%%%%%%%%%%%%%%%%%%%%%%
%%%%%%%%%%%%%%%%%%%%%%%%%%%%%%%%%

%%%%%%%%%%%%%%%%%%%%%%%%%%%%%%%%%
%%%%%%%%%%%%%%%%%%%%%%%%%%%%%%%%%
\end{document}